\documentclass[11pt]{article}

\usepackage{jheppub}
\usepackage{amsmath,amssymb}
\usepackage{calc}
\usepackage{placeins}
\usepackage{xspace}

 \usepackage{color}
 \definecolor{darkgreen}{rgb}{0,0.5,0}
 \definecolor{darkblue}{rgb}{0,0,0.7}
 \definecolor{darkred}{rgb}{0.5,0,0.0}
 \definecolor{darkorange}{rgb}{0.8,0.4,0.0}

\newcommand{\GeV}{\,\mathrm{GeV}}

\newcommand{\as}{\alpha_s}

\newcommand{\beq}{\begin{equation}}
\newcommand{\eeq}{\end{equation}}
\newcommand{\bea}{\begin{eqnarray}}
\newcommand{\eea}{\end{eqnarray}}
\newcommand{\bdm}{\begin{displaymath}}
\newcommand{\edm}{\end{displaymath}}

\def\as{\alpha_s}

\def\ord{{\cal O}}

\def\d{\partial}

\def \d{{\rm d} }
\def \d0 {D\O \;}

\def \yc {y_{\mathrm{cut}}}
\def \zc {z_{\mathrm{cut}}}
\def \d {\mathrm{d}}
\newcommand{\zcut}{z_{\text{cut}}}

\newcommand{\pt}{p_{t,\text{jet}}}
\newcommand{\ptg}{p_{t,\text{\text{mMDT}}}}
\newcommand{\lr}{\log \left(\dfrac{1}{\rho} \right)}

\newcommand{\herwig}{\textsf{Herwig}}
\newcommand{\pythia}{\textsf{Pythia}}

\newcommand{\fastjet}{\textsf{FastJet}}
\newcommand{\fjcontrib}{\textsf{fjcontrib}}
\newcommand{\eventtwo}{\textsf{EVENT2 }}
\newcommand{\nlo}{\textsf{NLOJet++}\xspace} 

\makeatletter
\g@addto@macro\bfseries{\boldmath}
\makeatother

\title{A study of jet mass distributions with grooming}

\author[a]{Simone Marzani,}
\author[b]{Lais Schunk,}
\author[b]{and Gregory Soyez}
\affiliation[a]{University at Buffalo, The State University of New York, Buffalo New York 14260-1500, USA}
\affiliation[b]{IPhT, CEA Saclay, CNRS UMR 3681, F-91191 Gif-Sur-Yvette, France}

\preprint{{\flushright}}


\abstract{ We perform a phenomenological study of the invariant mass
  distribution of hadronic jets produced in proton-proton collisions,
  in conjunction with a grooming algorithm. 
  In particular, we consider the modified MassDrop Tagger (mMDT), which corresponds to Soft Drop with angular exponent $\beta=0$.
  Our calculation, which is
  differential in both jet mass and jet transverse momentum, resums
  large logarithms of the jet mass, including the full dependence on the groomer's energy threshold $\zcut$,
  and it is matched to fixed-order QCD matrix elements at next-to-leading order.
  In order to account for non-perturbative contributions, originating
  from the hadronisation process and from the underlying event, we
  also include a phenomenological correction factor derived from Monte
  Carlo parton shower simulations.
  Furthermore, we consider two different possibilities for the jet
  transverse momentum: before or after grooming.
  We show that the former should be
    preferred for comparisons with upcoming experimental data essentially because the mMDT transverse momentum spectrum is not collinear safe,
 though the latter exhibits less sensitivity to underlying event and displays properties that may provide complementary information for probing non-perturbative effects.
}
\begin{document}

\maketitle
\section{Introduction}\label{sec:intro}
The CERN Large Hadron Collider has been running at an energy of 13~TeV
in the centre-of-mass frame, thus reaching energies far above the
electroweak scale.
Consequently, $Z/W^\pm$, Higgs, top quarks and any new particle with a
mass around the electroweak scale can be produced with a large boost,
causing their hadronic decays to become collimated so that they may be
reconstructed as a single
jet~\cite{Seymour:1993mx,Butterworth:2002tt}.
As a results, jet substructure is playing a central role during Run-II
of the LHC and its importance is only going to increase for future
runs, as well as at future higher-energy
colliders~\cite{Abdesselam:2010pt,Altheimer:2012mn,Altheimer:2013yza,Adams:2015hiv}.
For example, even though not confirmed in Run-II (see
e.g.~\cite{ATLAS:2016yqq}), an interesting excess in the invariant
mass distribution of two $W$ bosons was observed with Run-I
data~\cite{Aad:2015owa,Khachatryan:2014hpa}, relying on
jet-substructure techniques to isolate the signal from the QCD
background.

Jet substructure studies aim to better understand radiation patterns
in jets, in order to build efficient algorithms that can distinguish
signal jets from the QCD background. Examples include jet angularities~\cite{Berger:2003iw,Ellis:2010rwa},  energy-energy
correlation functions~\cite{Larkoski:2013eya,Moult:2016cvt}, and other jet shapes~\cite{Seymour:1997kj,Thaler:2010tr,Thaler:2011gf,Salam:2016yht,Dasgupta:2016ktv,Larkoski:2015kga,Larkoski:2014gra,Larkoski:2014zma} of
high-$p_t$ jets. Perhaps
the simplest example of such observables is the jet invariant
mass. Signal jets, which originate from the decay of a boosted massive
particle, are expected to have a mass in the region of that massive
state. QCD jets instead acquire mass through parton branching and
their mass is proportional to the jet transverse momentum. Thus, a cut
on the jet mass could be, in principle, a good discriminant. However,
many issues come into play and make this simple picture too
naive. First, the mass of QCD jets often appears to be in
  the same range as the signal jets. Then, radiation can leak outside the jet, altering both
signal and background. Moreover, hadron colliders are not clean
environments and there are many sources of additional,
non-perturbative, radiation that pollute the parton-level picture,
e.g.\ the underlying event, caused by secondary scattering in a
proton-proton collision, and pile-up, caused by multiple proton-proton
interactions.

For the above reasons, many substructure algorithms, often referred to
as ``groomers" and ``taggers", have been developed. Broadly
speaking, a grooming procedure takes a jet as an input and tries to
clean it up by removing constituents which, being at wide angle and
relatively soft, are likely to come from contamination, such as the
underlying event or pile-up. A tagging procedure instead focuses on
some kinematic variable that is able to distinguish signal from
background, such as, for instance, the energy sharing between two
prongs within the jet, and cuts on it.
Many of the most commonly used substructure algorithms such as the
MassDrop Tagger (MDT)~\cite{Butterworth:2008iy},
trimming~\cite{Krohn:2009th},
pruning~\cite{Ellis:2009su,Ellis:2009me}, or Soft
Drop~\cite{Larkoski:2014wba} perform both grooming and tagging, so a
clear distinction between the two is not always obvious.
These techniques have now been successfully tested and are currently
used in experimental analyses.

A quantitative understanding of groomed jet cross sections and
distributions is of paramount importance not only in order to devise
more efficient substructure algorithms but also in order to understand
their systematics, thus assessing their robustness. For instance, the
study of Refs.~\cite{Dasgupta:2013ihk,Dasgupta:2013via} revealed
unwanted features (kinks) in the mass distribution of background jets
with certain grooming algorithms, such as trimming and pruning, that
deteriorate the discrimination power at high $p_t$. Therefore, more
robust grooming techniques, with better theoretical properties, such
as the modified MassDrop Tagger (mMDT)~\cite{Dasgupta:2013ihk} and
Soft Drop~\cite{Larkoski:2014wba}, defined in Section~\ref{sec:mmdt},
were developed in order to overcome these issues.
A deeper understanding of these tools can be achieved by comparing
accurate theoretical predictions to data. On the experimental side,
one would like to have unfolded distributions of substructure
variables measured on QCD jets, as for instance in
Refs.~\cite{ATLAS:2012am,Chatrchyan:2013vbb}. On the theory side,
all-order calculations have been performed for a variety of
substructure variables with Soft Drop (or
mMDT)~\cite{Larkoski:2014wba}, such as the jet invariant mass, energy
correlations, the effective groomed radius and the prongs' momentum
sharing~\cite{Larkoski:2015lea}. More recently, using the framework of
Soft-Collinear Effective Theory (SCET), these calculations have
achieved the precision frontier, reaching next-to-next-to-leading
logarithmic accuracy (NNLL)~\cite{Frye:2016okc,Frye:2016aiz}, albeit in
some approximation, such as the small-$\zcut$ limit, as we will
discuss in what follows.
Furthermore, it has been shown that jet observables with grooming are
less sensitive to non-perturbative corrections than traditional
ones. This was expected in the case of contamination from the
underlying event and pile-up because groomers are indeed designed
 to remove soft radiation at large angle, which constitutes the
bulk of these contributions. Less obvious, but now understood from a
variety of Monte Carlo simulations as well as theoretical
considerations, is the reduction of the hadronisation
contribution. These properties contribute to make groomed distribution
even more amenable for comparisons between data and calculations in
perturbative QCD.

In this paper, we perform a phenomenological study of the jet mass
distribution with mMDT --- also
corresponding to Soft Drop with $\beta=0$ --- motivated by an upcoming
CMS measurement~\cite{CMS-coming}.
We consider jet mass distributions in several transverse momentum
bins. Our theoretical prediction accounts for the resummation of the
leading large logarithms of the ratio of the jet mass over the jet
transverse momentum and it is matched to fixed-order matrix elements
computed at next-to-leading order (NLO). While the accuracy of this
resummation is one logarithmic order less than the one presented in
Refs.~\cite{Frye:2016okc,Frye:2016aiz} in the case $\beta=0$,\footnote{See
    the discussion below Eq.~(\ref{sigma_log}) for our counting of the
    logarithmic accuracy.} we do lift the small-$\zcut$ approximation.
Crucially, working at finite $\zcut$ allows us to keep track of the
distinction between the jet transverse momentum before or after
grooming, henceforth $\pt$ and $\ptg$, respectively.
The two are, of course, equal at $\zcut = 0$.
We find that the use of $\ptg$ has several theoretical disadvantages
with respect to $\pt$. While the two resummations coincide as
$\zcut \to 0$, the $\ptg$ selection leads to a more involved perturbative
structure even at the leading nontrivial order. This difference stems
from a basic fact, namely while the ungroomed $\pt$ spectrum is an
Infra-Red and Collinear (IRC) safe quantity, the jet $\ptg$ spectrum
(with no additional cuts) is Sudakov safe~\cite{Larkoski:2013paa,
  Larkoski:2014wba,Larkoski:2015lea} but not IRC safe. Conversely, the
$\ptg$ spectrum is less sensitive to the underlying event than $\pt$
one and, arguably, more resilient to pile-up.
It is therefore interesting to explore both options in more details.

This paper is organised as follows. In Section~\ref{sec:mmdt} we
review definition and properties of Soft Drop and mMDT. Resummation
and matching of the mass distribution with $\pt$ are done in
Section~\ref{sec:ptjet}, followed by the case of $\ptg$ in
Section~\ref{sec:ptgroomed}. A Monte Carlo study of non-perturbative
corrections is presented in Section~\ref{sec:np-corrections}, while we
collect our final phenomenological predictions in
Section~\ref{sec:final}. Finally, we conclude in
Section~\ref{sec:conclusion}.

\section{A brief reminder of the grooming procedure}\label{sec:mmdt}
The Soft Drop grooming procedure~\cite{Larkoski:2014wba} takes a jet
with momentum $\pt$ and radius $R$. It re-clusters its constituents
using the Cambridge/Aachen (C/A) algorithm \cite{Dokshitzer:1997in,
  Wobisch:1998wt} and iteratively performs the following steps:
\begin{enumerate}
 \item it de-clusters the jet into 2 subjets $j \to j_1 + j_2$;
 \item it checks the condition 
\begin{equation}\label{eq:sd-condition}
\frac{\min (p_{t1} , p_{t2})}{p_{t1}+p_{t2}} > \zc \left(
  \frac{\theta_{12}}{R}\right)^\beta\,;
\end{equation}
\item if the jet passes the condition, the recursion stops; if not the
  softer subjet is removed and the algorithms goes back to step 1 with
  the hardest of the two subjets. 
 \end{enumerate}
 As previously anticipated, in this study we only consider
 $\beta=0$. In this case Soft Drop essentially reduces to the mMDT,
 albeit without any actual mass-drop condition. Moreover, while the
 original MDT algorithm imposed a cut on the ratio of angular distances
 to masses, a so-called $\yc$, the mMDT variant instead cuts on
 momentum fractions~\cite{Dasgupta:2013ihk} (see
 e.g. \cite{Dasgupta:2013ihk,Dasgupta:2016ktv} for a comparison
 between $\yc$ and $\zc$).

 From a theoretical point of view, Soft Drop has numerous advantages. 
 For instance, non-global
 logarithms~\cite{Dasgupta:2001sh,Dasgupta:2002bw}, which require
 sophisticated treatments, e.g.~\cite{Banfi:2002hw,Forshaw:2006fk,
   Forshaw:2008cq,DuranDelgado:2011tp,Weigert:2003mm,
   Hatta:2013iba,Schwartz:2014wha,
   Larkoski:2015zka,Larkoski:2016zzc,Neill:2016stq,
   Caron-Huot:2015bja, Becher:2015hka,Becher:2016mmh} and are often
 the bottle-neck of resummed calculations, are removed.
 Moreover, if we concentrate on mMDT, as we do here, the perturbative behaviour of observables such as the jet
 mass, which are double-logarithmic when measured on ungroomed
 jets, is changed into a single-logarithmic one because the soft-collinear region of phase-space is groomed away.
 Furthermore, the action of the groomer greatly reduces the impact of
 non-perturbative contributions, such as hadronisation, the underlying
 event and pile-up, extending the regime of validity of perturbation
 theory down to smaller values of the observables of interest. This
 opens up the possibility of performing sensible comparisons between
 data and first-principle theoretical predictions across a significant
 region of phase-space.
 
\section{Jet mass distributions with mMDT}\label{sec:ptjet}

Throughout this paper, we focus on the invariant mass of a mMDT jet
produced in proton-proton collisions with a centre-of-mass energy of
13 TeV.
Our selection cuts closely follow the ones of the upcoming CMS
measurement~\cite{CMS-coming}: jets are defined with the anti-$k_t$
algorithm~\cite{Cacciari:2008gp} with jet radius $R=0.8$.  Next, we
select the two hardest jets, $j_a$ and $j_b$, of the event and impose
the following conditions:
\begin{enumerate}
\item both jets must have $\pt > 200 \GeV$ and central rapidity, namely $|y| < 2.4$;
\item the transverse momenta of the jets must satisfy $|p_{ta} - p_{tb}| < 0.3 (p_{ta} + p_{tb})$, in order to select symmetric configurations;
\item the jets should be well-separated in azimuth, i.e. $\Delta \phi_{j_a, j_b} > \pi/2$.
\end{enumerate}
In practice, these cuts are intended to select dijet events. We note
however that the transverse momentum cut on the second jet results in
large perturbative corrections for the dijet cross-section which
render the mass distribution unstable in the first transverse momentum
bin. Imposing only a $p_t$ cut on the leading jet and the symmetry
condition would have been similarly efficient at selecting dijet
events, and would have improved the perturbative convergence.

For every jet that passes the above cuts, we apply the mMDT procedure
with $\zc=0.1$. We compute the (groomed) jet mass squared defined as
$m^2= \left( \sum_{i} p_i\right)^2,$ where the sum runs over all
particles in the groomed jet.  We also find useful to define the
dimensionless variable
\begin{equation}\label{jetrho}
\rho= \frac{m^2}{\pt^2 R^2}.
\end{equation}
We calculate the $\rho$ distribution in a given transverse momentum bin $p_{t1}< \pt< p_{t2}$:
\begin{equation}\label{sigmabin}
\frac{d \sigma}{d \rho}\left( \rho; \zc, p_{t1},p_{t2} \right)= \int_{p_{t1}}^{p_{t2}} d \pt \frac{d^2 \sigma}{d \pt d \rho}.
\end{equation}
We also define the normalised distribution as
\begin{equation}\label{sigmabin-norm}
\frac{d \tilde{\sigma}}{d \rho}\left( \rho; \zc, p_{t1},p_{t2} \right)=\frac{1}{\sigma_\text{bin}(p_{t1},p_{t2})}
\frac{d \sigma}{d \rho}\left( \rho; \zc, p_{t1},p_{t2} \right),
\end{equation}
where $\sigma_\text{bin}$ is the jet cross section in the transverse momentum bin under
consideration. 
We also explicitly consider the jet mass distribution
\begin{equation}\label{sigma-mass-bin}
\frac{d \sigma}{d m}\left( m; \zc, p_{t1},p_{t2} \right)= \int_{p_{t1}}^{p_{t2}} d \pt \frac{d^2 \sigma}{d \pt d m}=
\int_{p_{t1}}^{p_{t2}}d \pt \frac{2m}{\pt^2R^2}  \frac{d^2 \sigma}{d \pt d \rho},
\end{equation}
and the corresponding normalised version.
Furthermore, the quantity that is measured experimentally is the mass
distribution integrated over a set of mass bins $m_i<m<m_{i+1}$, which
is the observable we are going to explicitly show in our plots.
Note that in both Eq.~(\ref{jetrho}) and Eq.~(\ref{sigmabin}) $\pt$ is
the jet transverse momentum {\it before} grooming. We will consider
the alternative choice, namely the groomed transverse momentum $\ptg$
in Section~\ref{sec:ptgroomed}.

Analytic all-order calculations of jet shapes with grooming is a
rapidly developing field. In particular, the leading logarithmic
resummation of mMDT jet masses has been performed
in~\cite{Dasgupta:2013ihk} and resummation for Soft Drop observables,
i.e. for generic $\beta$, was
performed to NLL accuracy in~\cite{Larkoski:2014wba} and to NNLL
accuracy in~\cite{Frye:2016okc,Frye:2016aiz}.
All the logarithmic contributions in Soft Drop observables are of
collinear origin, while soft-emission at large angle can at most
contribute with logarithms of $\zc$. 
Thanks to this observation, the
resummed calculation can be done in the collinear limit and the
resulting structure is much simpler than the one that we encounter in
the resummation of the jet mass distributions without grooming, see
for instance~\cite{Dasgupta:2012hg,Jouttenus:2013hs,Chien:2012ur}. In
particular, soft radiation at large angle, which would give rise to a
nontrivial matrix structure in colour space, is groomed away: only
dipoles involving the measured jet are logarithmically enhanced and
require resummation, while initial-state radiation does not
contribute.
For the same reason, these observables are free of non-global logarithms.

At this stage, a word of caution about our counting of the
logarithmic accuracy is in order. While for a generic (non-zero)
$\beta$, the Soft Drop mass distribution is dominated by double
logarithms --- with LL accuracy resumming those double logarithms, NLL
accuracy including single-logarithms as well, etc... --- these double
logarithms are absent for mMDT (i.e. Soft Drop with $\beta=0$) in the region $\rho < \zc$:
\begin{equation}\label{sigma_log}
\rho \frac{d \tilde{\sigma}}{d \rho}\left( \rho;\zc \right)= \left[ \sum_{n=1}^\infty \sum_{m=1}^n c_{n,m}(\zc)\,\as^n \log^{m-1} \left(\frac{1}{ \rho }\right) + \mathcal{O}(\rho)\right],
\end{equation}
where the dependence on the transverse momentum bin is understood.
Single logarithmic terms in the jet mass are therefore formally the
leading contribution and will be referred to as LL in what follows.
The logarithmic counting of Refs.~\cite{Frye:2016okc,Frye:2016aiz}
differs from ours because it refers to the accuracy of the objects
that appear in the factorisation theorem. These functions are
separately double-logarithmic, even for $\beta=0$, and the
cancellation of the double logarithms only happens when they are
combined.\footnote{We would like to thank Andrew Larkoski for
  clarifying this point.}
In our counting, the NLL \cite{Larkoski:2014wba} and
NNLL~\cite{Frye:2016okc,Frye:2016aiz} results obtained for a generic $\beta$, actually correspond respectively to 
LL and NLL accuracy, in the small $\zc$ limit, for mMDT.
Thus, the state-of-the art evaluation of Eq.~(\ref{sigma_log}) accounts for all the coefficients
$\tilde{c}_{n,n}(\zc)$ and $\tilde{c}_{n,n-1}(\zc)$, where
\begin{equation}\label{c_coeff}
\lim_{\zc \to 0}c_{n,m}(\zc)=\tilde{c}_{n,m}(\zc)+ \mathcal{O}(\zc).
\end{equation}
For phenomenology, one typically uses $\zc \simeq 0.1$, so it is
important to investigate the size of finite $\zc$ corrections. In this
study we restrict ourselves to LL accuracy, while maintaining for the
full $\zc$ dependence, i.e.\ we fully account for all coefficients
$c_{n,n}(\zc)$.

Finally, in the region $\rho > \zc$ grooming is not active and we recover the
traditional jet mass result~\cite{Dasgupta:2013ihk}. In this region we are going to
perform a less sophisticated calculation which resums the double
logarithms and those single logarithmic contributions of collinear
origin. We find this procedure acceptable because in this region
$\rho \sim \zc$ and we expect these contributions to be less important
than the fixed-order corrections, which we include at NLO.

\subsection{Resummation at finite $\zc$}\label{sec:llresum}
The resummation of the mMDT mass distribution at finite $\zc$ was
outlined in Ref.~\cite{Dasgupta:2013ihk} in the fixed-coupling
limit.\footnote{More precisely, the resummation of
  Ref.~\cite{Dasgupta:2013ihk} was performed in case of a $\yc$, but
  its modification to a $\zc$ is straightforward.} The major
complication with respect to the small-$\zc$ limit has to do with the
flavour structure. Let us consider for instance a $q \to q g$ splitting
which does not satisfy the mMDT condition. There is an
$\mathcal{O}(\zc)$ probability for the gluon to be harder than the
quark. In such a case, the declustering sequence would follow the
gluon branch rather than the quark, resulting into a nontrivial mixing
between quarks and gluons.
The resummed distribution therefore acquires a matrix
structure in flavour space~\cite{Dasgupta:2013ihk}
\begin{equation}
\label{eq:matrix}
\rho \dfrac{d^2 \sigma}{d \pt d \rho} = (R'_{q} \; \;  R'_{g}) \exp
 \left( \begin{array}{cc}
-R_{q}-R_{q \to g} & R_{g \to q}  \\
R_{q \to g} & -R_{g} - R_{g \to q} \end{array} \right)
 \left( \begin{array}{c}
\sigma_q  \\
\sigma_g \end{array} \right),
\end{equation}
where $\sigma_{q(g)}$ is Born-level cross section for a quark (gluon)
with transverse momentum $\pt$ and $R'_{q(g)}= \partial_L R_{q(g)}$,
with $L=\log(1/\rho)$.
As previously discussed, because we are dealing with a Soft Drop
observable, the radiators $R_i$ can be computed in the collinear
limit. Denoting by $\theta$ the emission angle (in units of the jet
radius $R$) with respect to the hard momentum and with $z$ the
momentum fraction, we have~\footnote{ For simplicity, we introduce the
  following notation for the Heaviside step function:
  $\Theta\left( a>b\right)\equiv \Theta\left( a-b\right)$,
  $\Theta\left( a<b\right)\equiv \Theta\left( b-a\right)$, and
  $\Theta\left( a<b<c\right)\equiv \Theta\left(b-a \right)
  \Theta\left(c-b \right)$.  }
\begin{subequations} \label{eqn:sudakov-integral-representation}
\begin{align}
R_q &= C_F \int_{0}^{1} \dfrac{d \theta^2}{\theta^2} \int_0^1 d z \; p_{gq} (z) \dfrac{\alpha_s (z \theta \pt R)}{\pi}  \Theta \left(\zc<z<1-\zc \right) \Theta(z \theta^2 > \rho),\label{eq:Rq-full-expr}\\
R_g &=  C_A \int_{0}^{1} \dfrac{d \theta^2}{\theta^2}\int_0^1 d z \;
      p_{xg} (z) \dfrac{\alpha_s (z \theta \pt R)}{\pi} \Theta \left(
      \zc < z < 1- \zc \right)  \Theta(z \theta^2 > \rho)\label{eq:Rg-full-expr},\\
R_{q \to g} &\:=\: C_F \int_{0}^{1} \dfrac{d \theta^2}{\theta^2} \int_0^1 d z \; p_{gq} (z) \dfrac{\alpha_s (z \theta \pt R)}{\pi} \Theta \left(1-z < \zc \right) \Theta(z \theta^2 > \rho),\label{eq:Rq2g-full-expr}\\
R_{g \to q} &=  T_R n_f \int_{0}^{1} \dfrac{d \theta^2}{\theta^2} \int_0^1 d z \; p_{qg} (z) \dfrac{\alpha_s (z \theta \pt R)}{\pi}  \left[ \Theta \left(1-z< \zc \right) + \Theta \left( z < \zc\right)\right] \Theta(z \theta^2 > \rho),\label{eq:Rg2q-full-expr}
\end{align}
\end{subequations}
where $C_A = 3$, $C_F = 4/3$, $T_R=1/2$, $n_f$ is the number of active
quark flavours and $p_{ab}(z)$ are the splitting functions given in
Appendix~\ref{app:calc}. 

At the LL accuracy we are working at, the above expressions can be further
simplified. Besides the strict leading-logarithmic terms in $\rho$, it
is trivial to also include the double-logarithmic terms in $\zc$
and this allows for a more transparent treatment of the transition
point at $\rho=\zc$.
In that context, it is helpful to separate
Eq.~(\ref{eqn:sudakov-integral-representation}) in a contribution
$\mathcal{R}_i$, coming from the $1/z$ part of the splitting
function that includes the logarithmic and constant terms in $\zc$,
and a remainder which contains the corrections power-suppressed in
$\zc$. Later, this will make it easy to study the size of the
finite-$\zc$ corrections.
For these contributions, we neglect the $z$ factor in
the argument of $\alpha_s$ and in the constraint $z\theta^2>\rho$.
The details of our calculation are given in Appendix~\ref{app:calc}
and, our final result reads
\begin{subequations}\label{eq:sudakov}
  \begin{align}
  R_q &= C_F\mathcal{R}_q(\rho;\zc)\, \Theta(\rho<e^{B_q}) + C_F\,\mathcal{I}(\rho;\zc)\, \pi_q(\zc)\, \Theta(\rho<\zc),\\
  R_g &= C_A\mathcal{R}_g(\rho;\zc)\, \Theta(\rho<e^{B_g}) + C_A\,\mathcal{I}(\rho;\zc)\, \pi_g(\zc)\, \Theta(\rho<\zc),\\
  R_{q \to g} &= C_F\,\mathcal{I}(\rho;\zc)\, \pi_{q\to g}(\zc)\, \Theta(\rho<\zc),\\
  R_{q \to g} &= n_fT_R\,\mathcal{I}(\rho;\zc)\, \pi_{g\to q}(\zc)\, \Theta(\rho<\zc),
  \end{align}
\end{subequations}
where we have introduced
\begin{subequations}\label{eq:Rsoft-and-I}
\begin{align}
  \mathcal{R}_i(\rho;\zc)
  & = \frac{1}{2\pi\alpha_s\beta_0^2}\Big[
   W\big(1+2\alpha_s\beta_0B_i\big)
   -W\big(1+2\alpha_s\beta_0\log(z_m)\big)\\
& \phantom{ = \frac{1}{2\pi\alpha_s\beta_0^2}}
   +2 W\big(1+\alpha_s\beta_0\log(\rho z_m)\big)
   -2 W\big(1+\alpha_s\beta_0(\log(\rho)+B_i)\big)\Big],\nonumber\\
    \mathcal{I}(\rho;\zc)
  &= \int_\rho^{\zc} \frac{dx}{x}\frac{\alpha_s(xp_tR)}{\pi} =
    \frac{1}{\pi \beta_0}\log\bigg(\frac{1+\alpha_s\beta_0\log(\zc)}{1+\alpha_s\beta_0\log(\rho)} \bigg),
\end{align}
\end{subequations}
with $W(x)=x\log(x)$, $z_m=\text{max}(\zc,\rho)$, $B_q=-\frac{3}{4}$,
$B_g=-\frac{11}{12}+ \frac{n_fT_R}{3 C_A}$ and
\begin{subequations}\label{eq:splitting-integrals}
\begin{align}
  \pi_q(\zc) & = \log(1-\zc) +
               \frac{3\zc}{2} ,\\
  \pi_g(\zc) & = 
               \log(1-\zc)+2\zc-\frac{\zc^2}{2}+\frac{\zc^3}{3}
               -\frac{n_fT_R}{C_A}\Big(\zc-\zc^2+\frac{2\zc^3}{3}\Big),\\
  \pi_{q\to g}(\zc) & = -\log(1-\zc)-\frac{\zc}{2}-\frac{\zc^2}{4},\\
  \pi_{g\to q}(\zc) & =  \zc-\zc^2+\frac{2\zc^3}{3}.
\end{align}
\end{subequations}
We note that the diagonal radiators vanish for $\rho=\exp(B_i)$ and,
since $B_q$ is (slightly) larger than $B_g$, this produces
distributions with an end-point at $\rho=\exp(B_q)$.
Furthermore, the appearance of $z_m=\text{max}(\zc,\rho)$ reproduces
the transition point at $\rho=\zc$, when the mMDT becomes active. We
show explicitly below that it corresponds to a transition between a
plain jet mass behaviour at large mass and a single-logarithmic
behaviour at low mass.

To gain some insight in this direction, it is helpful to consider the
limit of these expressions in a fixed-order approximation, where we
find
\begin{subequations} \label{eq:fixed-coupling-limit}
\begin{align}
  \mathcal{R}_i^{\text{(f.c.)}}(\rho;\zc)
  & = \frac{\alpha_s}{2\pi}\Big[\big(\log(\rho)-B_i\big)^2 - \log^2(z_m/\rho)\Big],\\
    \mathcal{I}^{\text{(f.c.)}}(\rho;\zc)
  &= \int_\rho^{\zc} \frac{dx}{x}\frac{\alpha_s(xp_t R)}{\pi} =
   \frac{ \alpha_s}{\pi}\log\Big(\frac{\zc}{\rho}\Big).
\end{align}
\end{subequations}
This clearly shows that the distribution is double-logarithmic for
$\rho>\zc$ (where $z_m=\rho$), and becomes single-logarithmic for
$\rho<\zc$ (where $z_m=\zc$). In the latter case, we also see that the
finite-$\zc$ corrections, proportional to $\mathcal{I}$ are entering at
the same order as the small-$\zc$ contributions, that is at the
leading-logarithmic accuracy.
Thus, these contributions must be included to formally obtain the full LL result.

In order to assess perturbative uncertainties we follow a standard
procedure.
We vary the factorisation scale (in the Born-level cross-sections
$\sigma_q$ and $\sigma_g$) and the renormalisation scale (both in the
resummation formula and in the Born-level cross-sections) by
a factor of two around the hard scale $\pt R$, keeping the ratio of scales never larger than 2 or smaller than 1/2, i.e.\ we employ a canonical 7-point scale variation~\cite{Cacciari:2003fi}.
We also introduce a resummation scale $\mu_Q$, which we use
to rescale the argument of the logarithms we are resumming
$L=\log \frac{\pt R}{\mu_\text{Q} \rho}$. We use variations of $\mu_Q$
by a factor of 2 around the hard scale $\pt R$ to assess the size of
logarithmic contributions beyond our accuracy.

\subsection{Fixed-order calculations and matching prescription}
\label{sec:match-ptjet} 
The resummed jet mass spectrum discussed in the previous section is reliable in the $\rho\ll1$ region, where the distribution is dominated by collinear splittings. In order to accurately describe the $\rho \sim 1$ region we have to resort to fixed-order computations. Ultimately, we will match the two calculations yielding theoretical predictions which are accurate at both small and large $\rho$, as discussed in the following. 

All our fixed-order predictions are obtained using the public code
\nlo~\cite{Catani:1996vz, Nagy:2003tz} together with the parton
distribution set CT14~\cite{Dulat:2015mca} at NLO. Jets are then
clustered with the anti-$k_t$ algorithm as implemented in
\fastjet~\cite{Cacciari:2005hq,Cacciari:2011ma} and we use the
implementation of mMDT in \fjcontrib~\cite{fjcontrib}. Jet mass
distributions are obtained by considering $2 \to 3$ partonic processes
at LO and NLO. Moreover, we also use \nlo to calculate the bin cross
section $\sigma_\text{bin}$, see Eq.~(\ref{sigmabin-norm}), and the
quark and gluon cross sections, $\sigma_q$ and $\sigma_g$
respectively. In order to estimate the theoretical uncertainty, we vary
renormalisation and factorisation scales around the central value
$\mu_\text{R}=\mu_\text{F}= \pt R$, with the 7-point method.

We are now ready to match the resummed and the fixed-order
calculations. Before discussing different matching schemes, we address
the issue of the end-point of the distribution at large $\rho$. It is
not difficult to show, see e.g.~\cite{Dasgupta:2012hg}, that the LO
distribution has an end-point at
$\rho_{\max,\text{LO}}=\frac{1}{4}+\mathcal{O}\left( R^2\right)$. At
NLO up to three partons can be reconstructed in a single jet, leading
to $\rho_{\max,\text{NLO}}=\frac{25}{64}+\mathcal{O}\left( R^2\right)$
(see Appendix~\ref{app:rhomax} for details).
On the other hand, our resummed calculation has an end-point at
$\rho=\exp(B_q)$, see Eq.~(\ref{eq:sudakov}).
It is clearly desirable to match curves with the same end-point,
therefore we modify the argument of the logarithms in the resummation
in such a way that the resummed distribution has the same end-point as
the fixed-order it is matched to~\cite{Catani:1992ua}
\begin{equation}\label{end_point}
\lr  \to \log \left( \dfrac{1}{\rho} -\dfrac{1}{{\rho_{\max,i}}} +e^{-B_q} \right),
\end{equation}
where for $R=0.8$ the end-points are found to be
$\rho_{\max,\text{LO}} = 0.279303$ and
$\rho_{\max,\text{NLO}} = 0.44974$ (see Appendix~\ref{app:rhomax}).

The combination of resummed and fixed-order results comes with a
certain degree of ambiguity. Different matching schemes must produce
resummed and matched distributions, LO+LL and NLO+LL, at the quoted
accuracy but they can differ for terms that are subleading in both
logarithmic and fixed-order counting. The simplest matching scheme is
the additive one, which consists of adding the two results while
removing double counting. This scheme suffers from two issues.
Firstly, when matching to NLO fixed-order results, our LL
calculation only includes the leading $\alpha_s^2\log(1/\rho)$
contribution and misses the constant $\alpha_s^2$ term, so an additive
matching would tend to a constant at small $\rho$ which is not
physically correct.
Secondly, even at LO, matching with our LL calculation requires a
precise numerical calculation of the small-$\rho$ tail, which can be
delicate to reach in the fixed-order calculation.
Therefore, we have decided to employ an alternative matching scheme, namely multiplicative matching.
We discuss it in some detail for the NLO+LL case and then recover from
it the simpler LO+LL. Naively, multiplicative matching can be defined
as
\begin{equation}\label{matching_def}
\sigma_{\mathrm{NLO+LL},\text{naive}}^{(m)} = \dfrac{\sigma_\mathrm{LL}^{(m)} \; \sigma_\mathrm{NLO}^{(m)}}{\sigma_\mathrm{LL,NLO}^{(m)}},
\end{equation}
where, to keep the notation compact, $\sigma_X^{(m)}$ indicates the jet
mass differential distribution computed at accuracy $X$, i.e.\
$\sigma_X^{(m)}\equiv \frac{d \sigma_X}{d m}$.
This construction applies both to the normalised and unnormalised
distributions.

Equation~(\ref{matching_def}) is however not ideal either because at
NLO accuracy, the fixed-order cross-section turns negative at small
mass. Asymptotically both $\sigma_\mathrm{NLO}^{(m)}$ and
$\sigma_\mathrm{LL,NLO}^{(m)}$ would be negative and their ratio would
tend to 1 but there is a region where they would be close to zero and
where Eq.~(\ref{matching_def}) would therefore be unreliable.
To fix this issue, we can write the fixed-order distribution
explicitly as
\begin{align}\label{expansion_FO}
\sigma_\mathrm{NLO}^{(m)}  &= \sigma_\mathrm{LO}^{(m)}+ \as \delta_\mathrm{NLO}^{(m)},
\end{align}
while the expansion of the resummation to second order is
\begin{align}\label{expansion_resum}
\sigma_\mathrm{LL,NLO}^{(m)} &= \sigma_\mathrm{LL,LO}^{(m)} + \as \delta_\mathrm{LL,NLO}^{(m)}.
\end{align}
We can then substitute Eq.~(\ref{expansion_FO}) and
(\ref{expansion_resum}) into Eq.~(\ref{matching_def}) and expand to
the desired accuracy, to obtain
\begin{align}\label{match1}
\sigma_\mathrm{NLO+LL}^{(m)}&= \sigma_\mathrm{LL}^{(m)}
\left[\frac{\sigma_\mathrm{LO}^{(m)}}{\sigma_\mathrm{LL,LO}^{(m)}}+\as \left(\frac{\delta_\mathrm{NLO}^{(m)}}{\sigma_\mathrm{LL,LO}^{(m)}} -
\sigma_\mathrm{LO}^{(m)}\frac{\delta_\mathrm{LL,NLO}^{(m)} }{{\sigma_\mathrm{LL,LO}^{(m)}}^2}
 \right) 
\right].
\end{align}
This is the expression we use in order to obtain our matched results.
The LO+LL results can be easily deduced from the above expression by
simply dropping the $\mathcal{O}(\as)$ correction in brackets,
in which case the expression corresponds to what would have been
obtained with a naive multiplicative matching.
We can also define alternative matching schemes. For instance, we can work with cumulative distributions
\begin{equation}\label{cumulative-def}
\Sigma_X(m) = \int_0^m d m' \, \frac{d \tilde \sigma_X}{d m'} =1 +\as \Sigma_X^{(1)}+ \as^2 \Sigma_X^{(2)}+ \mathcal{O}\left( \as^3\right),
\end{equation}
and employ the so-called log-$R$ matching~\cite{Catani:1992ua}, which combines together the logarithm of the cumulative distributions. This results in
\begin{align} \label{logRmatching}
\Sigma_\text{NLO+LL}^{\text{log-}R}&= \Sigma_\text{LL}  \exp \left[ \as \left( \Sigma^{(1)}- \Sigma^{(1)}_\text{LL}\right)+ \as^2 \left( \Sigma^{(2)}- \Sigma^{(2)}_\text{LL}\right)- \frac{\as^2}{2} \left( {\Sigma^{(1)}}^2- {\Sigma^{(1)}_\text{LL}}^2\right)\right].
\end{align}
A comparison between the different matching schemes will be discussed in the following.

\subsection{Perturbative results}\label{ptjet:results}

\begin{figure}
  \centering
  \includegraphics[width=0.495\textwidth,page=1]{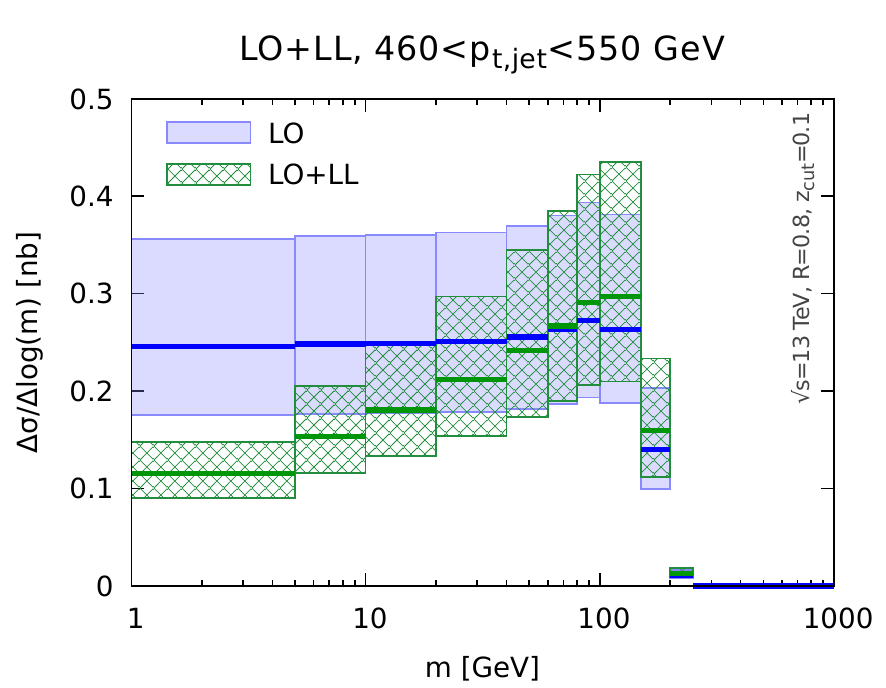}
  \includegraphics[width=0.495\textwidth,page=2]{figs-paper/ptjet-matched-LO.pdf}
  \includegraphics[width=0.495\textwidth,page=1]{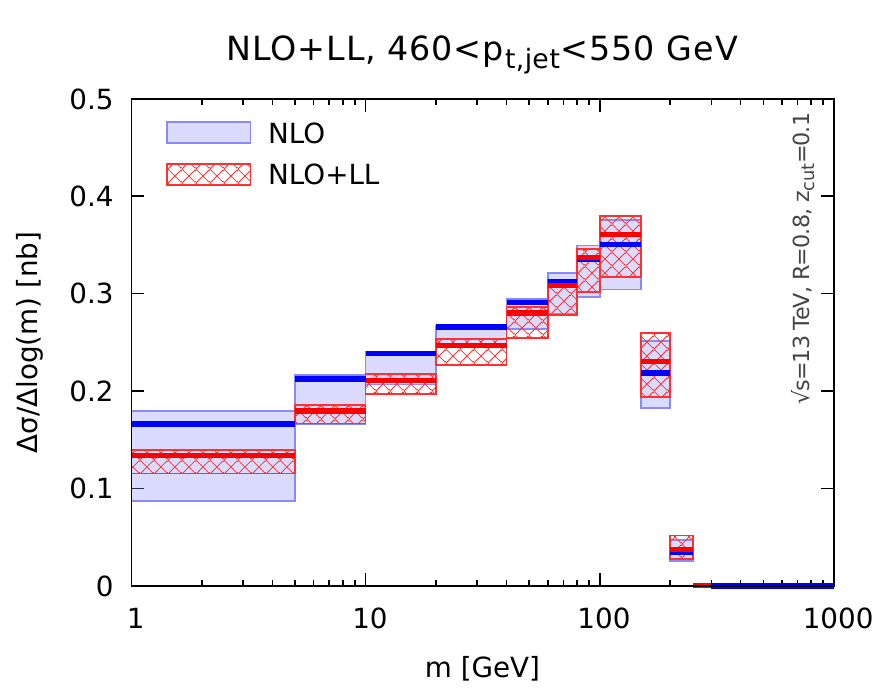}
  \includegraphics[width=0.495\textwidth,page=2]{figs-paper/ptjet-matched-NLO.pdf}
  \caption{In this figure we show the resummed and matched jet mass distribution in the $460< \pt< 550$~GeV transverse momentum bin (on the left), and in the $\pt>1300$~GeV bin (on the right). The top panels show LO+LL, while the bottom panels show NLO+LL.}
  \label{fig:distr}
\end{figure}

\begin{figure}[t]
  \centering
  \includegraphics[width=0.495\textwidth,page=1]{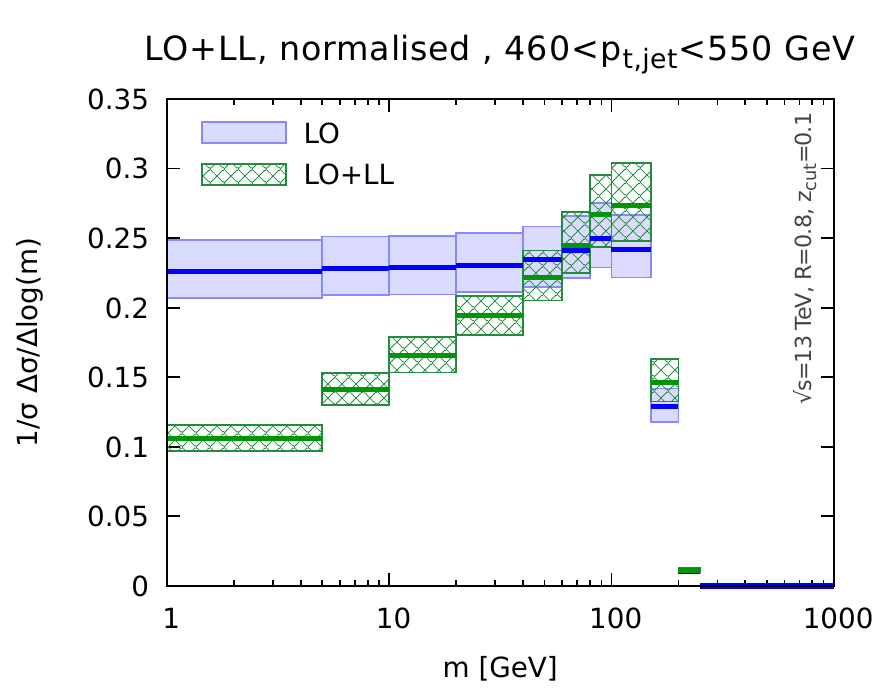}
  \includegraphics[width=0.495\textwidth,page=2]{figs-paper/ptjet-matched-LO-norm.pdf}
  \includegraphics[width=0.495\textwidth,page=1]{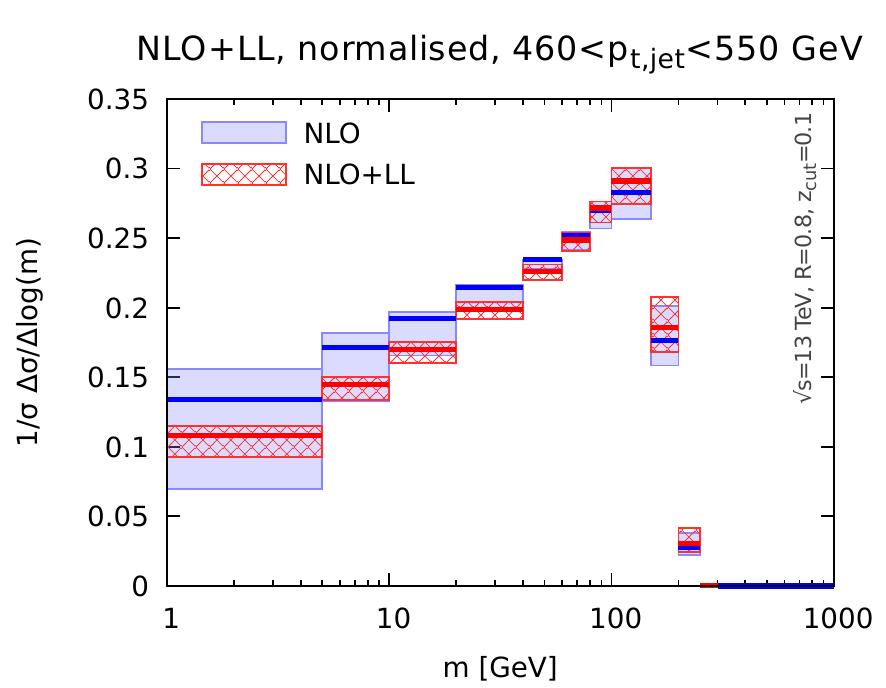}
  \includegraphics[width=0.495\textwidth,page=2]{figs-paper/ptjet-matched-NLO-norm.pdf}
  \caption{Same as in Fig.~\ref{fig:distr} but for the normalised distribution.}
  \label{fig:distr-norm}
\end{figure}

We now present our results for the resummed and matched jet mass distribution. We pick two representative bins in transverse momentum, namely $460< \pt< 550$~GeV and $\pt>1300$~GeV.
In Fig.~\ref{fig:distr} we show the mass distribution in logarithmic
bins of the mass:\footnote{The binned distribution is computed using
  Eq.~(\ref{sigma-mass-bin}). For a given $\pt$ we thus need
  to integrate $\rho {d^2 \sigma}/(d \pt d \rho)$ over a range in
  $\rho$. In practice, this can be written as a difference between the
  cumulative $\rho$ distribution taken at the bin edges,
  which, for the resummed results, is obtained by removing the
  $(R'_{q} \; \; R'_{g})$ pre-factor in Eq.~(\ref{eq:matrix}).}
\begin{equation}
\frac{\Delta \sigma}{\Delta \log m}\equiv \frac{m_{i+1}-m_{i}}{\log \left(m_{i+1}/m_{i} \right)} \frac{\Delta \sigma}{\Delta m},
\end{equation}
where $m_{i+1}$ and $m_i$ are, respectively, the upper and lower edge
of each mass bin.
Blue lines with a solid band represent distributions obtained with
fixed-order calculations and their uncertainty, while green or red
curves with a hatched band are for resummed and matched results obtained using Eq.~(\ref{match1}). 
We estimate the theoretical uncertainty on the matched result by taking the envelope of all the curves obtained by varying the arbitrary scales ($\mu_\text{R},\mu_\text{F}, \mu_\text{Q}$) which enter the fixed-order and resummed calculations, as previously detailed.
At the top we compare leading order distributions to LO+LL results, while at the bottom we show the NLO curve compared to NLO+LL. 
The plots on the left are for the lower-$\pt$ bin, while the ones on the right are for the boosted bin. 
We can see that the normalisation uncertainty is rather large especially when we consider LO distributions.  
Therefore, it is also interesting to look at normalised distributions,
with the normalisation taken to be the jet cross-section in the
relevant transverse momentum bin calculated at LO and NLO,
respectively for the LO(+LL) and NLO(+LL) results. We
show our results for the normalised distributions~in~Fig.~\ref{fig:distr-norm}.

\begin{figure}[t]
  \centering
   \includegraphics[width=0.495\textwidth,page=1]{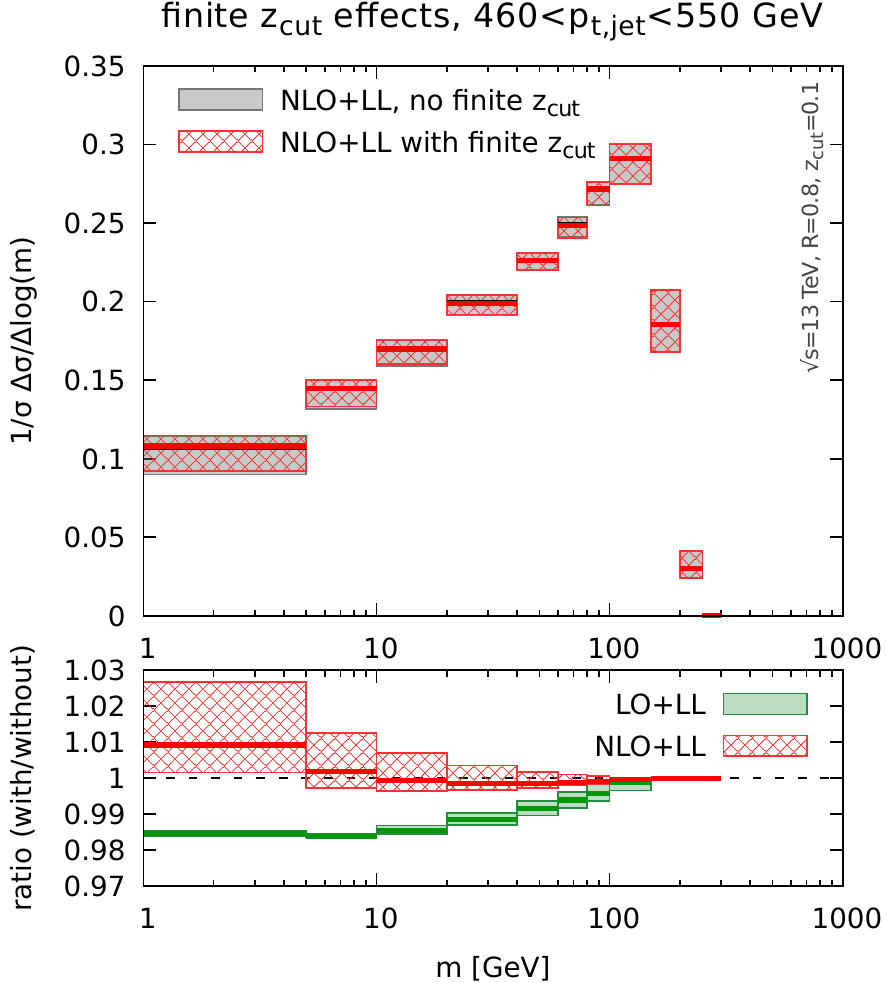}
   \includegraphics[width=0.495\textwidth,page=2]{figs-paper/ptjet-matched-finitez.pdf}
  \caption{Comparison between the resummed and matched calculation
    with finite $\zc$ (red) and the result with the resummation
    computed in the $\zc\to 0$ limit. The ratio plots at the bottom
    show that for $\zc=0.1$ this type of correction is very
    small.}
  \label{fig:finitezc}
\end{figure}

Since the state-of-the-art NLL
studies~\cite{Frye:2016okc,Frye:2016aiz} neglect the finite $\zc$
corrections, it is interesting to check their importance. In
Fig.~\ref{fig:finitezc} we compare the resummed and matched NLO+LL
normalised distribution, in red, to an approximation in which the
resummation is performed in the $\zc \to 0$ limit, in grey, for two
different transverse momentum bins.
From the top plots we can already see that, for $\zc=0.1$, these
effects are small and the two curves fall well within each other's
uncertainty bands.
Looking at the bottom plots we can see that these effects are at most a
couple of percent at NLO+LL (red curves).
For comparison, we also show, in green, the same ratio in the case of
the LO+LL result.
Note that the bands in the ratio plots represent the uncertainty on
the effect, not the overall uncertainty which is of the order of 10\%,
as can be seen from the top plots.
We also note that for a pure LL calculation, finite $\zcut$ effects
are found to be of the order of $2\%$, rising to about $5\%$ for
$\zcut =0.2$.
These findings justify the approximation of
Refs.~\cite{Frye:2016okc,Frye:2016aiz}, which achieved
higher-logarithmic accuracy but in the small-$\zc$ limit. We will see
in the next section that the situation radically changes when consider
bins in $\ptg$.

\begin{figure}[t]
  \centering
   \includegraphics[width=0.495\textwidth,page=1]{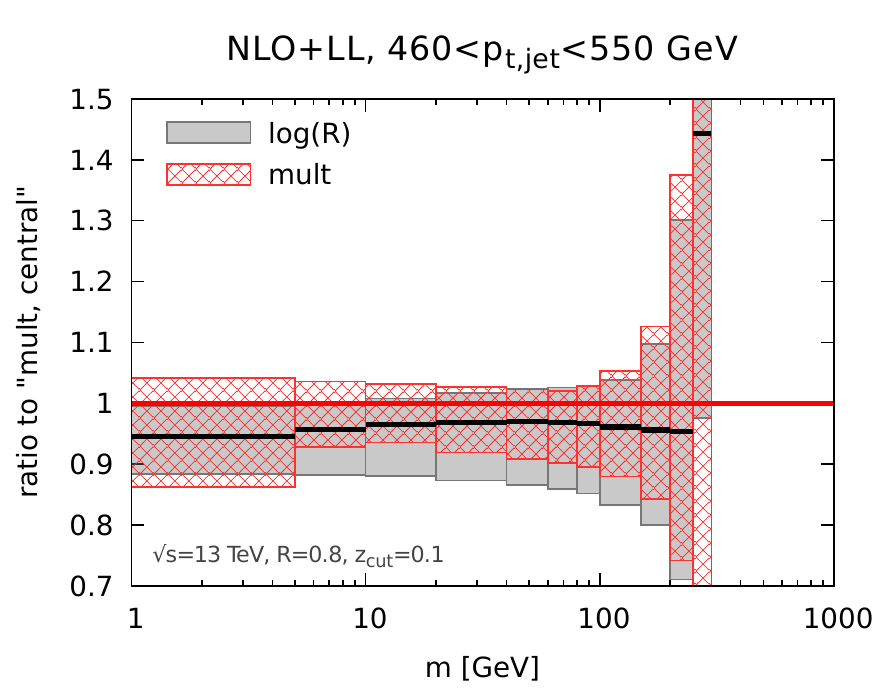}
   \includegraphics[width=0.495\textwidth,page=2]{figs-paper/ptjet-matching.pdf}
  \caption{Comparison of the jet mass distribution in two different matching schemes, the multiplicative one Eq.~(\ref{match1}) and the log-$R$ one Eq.~(\ref{logRmatching}).}
  \label{fig:logRmatching}
\end{figure}

Finally, in Fig.~\ref{fig:logRmatching} we compare two different matching schemes. In particular, we plot the ratio between the NLO+LL distribution obtained with log-$R$ matching Eq.~(\ref{logRmatching}) to the one obtained with multiplicative matching Eq.~(\ref{match1}), with their respective perturbative uncertainties. We see that the two results are in good agreement and they fall within each other's scale variation bands.

\FloatBarrier

 \section{Jet mass distributions with mMDT using $\ptg$}\label{sec:ptgroomed}
 
We now consider the alternative option where the mMDT jet mass is
measured in bins of $\ptg$ rather than $\pt$.
We begin our discussion pointing out a known but perhaps
under-appreciated fact: the transverse momentum distribution
$\frac{d \sigma}{d \ptg}$ is not IRC safe, see
e.g.~\cite{Larkoski:2014wba}.
We then proceed, as before, by discussing our calculation for the jet
mass distribution in bins of $\ptg$.

\subsection{Collinear unsafety  (but Sudakov safety) of $\ptg$}\label{sec:ptmmdt-safety}
The mMDT groomer only imposes a cut on the transverse momentum fraction
$z$. Therefore, real emissions below $\zc$ are groomed away without any
constraint on the emission angle, resulting in collinear
singularities that do not cancel against the corresponding virtual corrections.
Thus, the $\ptg$ distribution is IRC unsafe and it cannot be computed
order-by-order in the strong coupling $\as$, producing a divergence
even at the level of the first emission. However, this observable
still enjoys the property of Sudakov safety~\cite{Larkoski:2013paa,
  Larkoski:2014wba,Larkoski:2015lea} and it is therefore calculable
provided we perform an all-order computation.
We note that the situation is instead different if one considers Soft
Drop with $\beta>0$, which does regulate the collinear region.

One way to explicitly show the IRC unsafety of the $\ptg$ distribution
is to study fixed-order distributions in $e^+ e^-$ collisions using
the program \eventtwo~\cite{Catani:1996jh,Catani:1996vz}, for which we
can easily control the infrared cut-off scale.
In practice, we simulate events at Born level and at
${\cal O}(\alpha_s)$, including both real emissions and virtual
corrections. We cluster the full event with the $e^+e^-$ version of
the anti-$k_t$ algorithm with radius $R=0.4$ and select jets with an
energy larger than $0.95\sqrt{s}$/2, with $\sqrt{s}=1$~TeV.
We note that, at this order in perturbation theory, jets
have either one or two constituents.
We then run the following $e^+e^-$ version of mMDT: jets with one
constituent are kept untouched, and for jets with two constituents we
either keep them intact if $\text{min}(E_1,E_2)>\zc E_{\text{jet}}$, or
only keep the most energetic particle otherwise. We use $\zc=0.1$. We
consider the jet cross section for $E> 0.95\sqrt{s}/2$ before and
after applying mMDT.
At Born level, jets after the mMDT procedure are identical to the
ungroomed jets. At ${\cal O}(\alpha_s)$, for an initial jet with an
energy above the cut-off, the mMDT jet energy can drop below the
cut-off because of a collinear real emission inside the jet that does
not pass the mMDT condition. This cannot happen for virtual
corrections and so we do expect a leftover singularity.

\begin{figure}[t]
  \centering
  \includegraphics[width=0.495\textwidth]{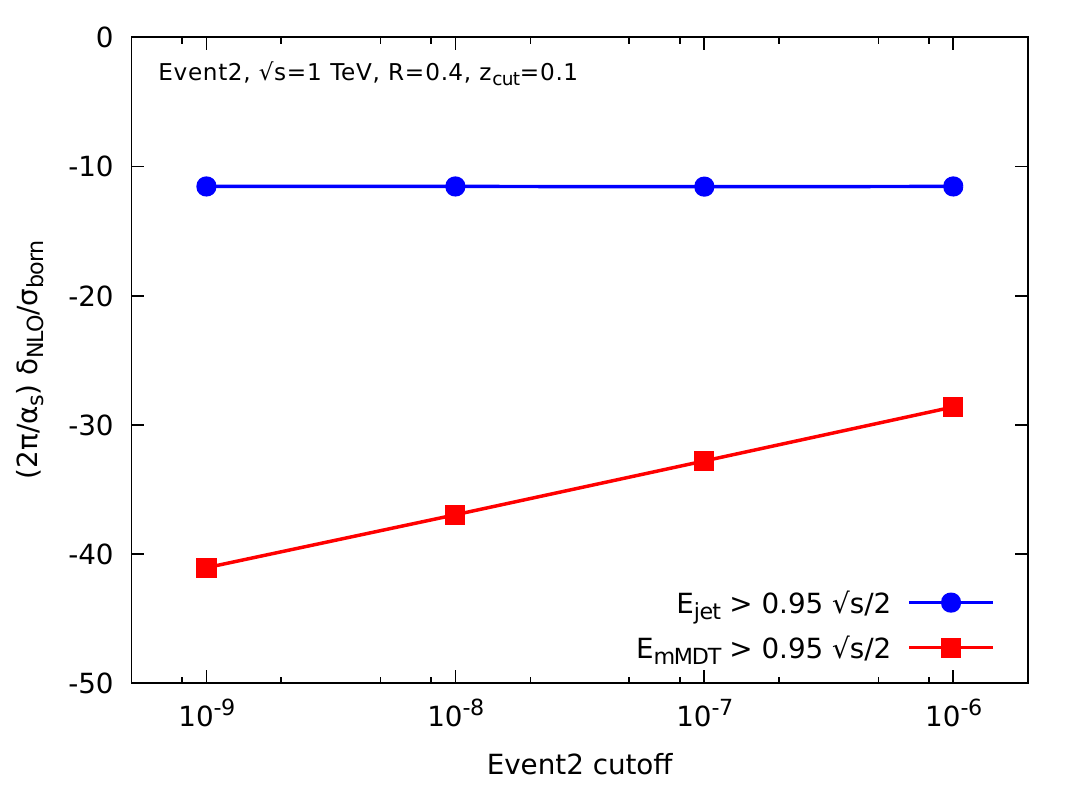}
  \caption{Dependence of the jet cross-section before and after
    applying mMDT, as a function of the infrared cut-off used in
    \eventtwo. The cross-section before grooming is stable but the one
    after grooming diverges logarithmically, thus making IRC unsafety apparent.}\label{fig:event2-cutoff-dependence}
\end{figure}

In numerical codes, both the real and the virtual terms are simulated
down to an infrared cut-off so that the numerical result is always
finite. When lowering the infrared cut-off parameter the cross section
for the ungroomed case is expected to remain constant (modulo small power
corrections), while the cross section for mMDT jets is expected to
have a residual logarithmic dependence on the cut-off as a consequence
of the IRC unsafety.
Fig.~\ref{fig:event2-cutoff-dependence} shows the results of our
simulations when varying the infra-red cut-off used in \eventtwo. We
indeed clearly see a constant behaviour for the (IRC safe) inclusive
cross-section and a logarithmically diverging behaviour for the (IRC
unsafe) cross-section after the mMDT procedure.

\begin{figure}[t]
  \centering
  \includegraphics[width=0.495\textwidth,page=1]{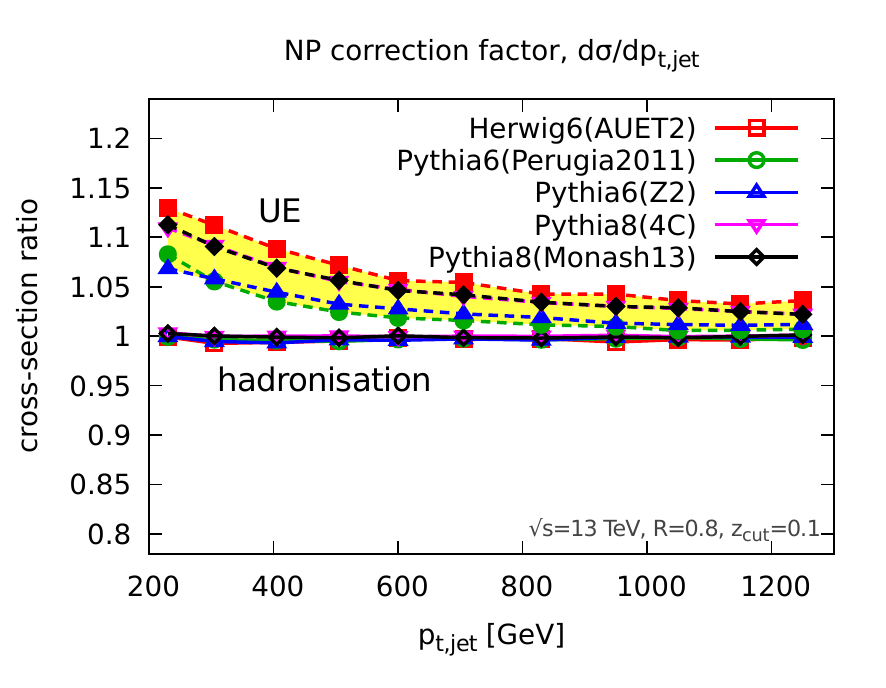}
  \includegraphics[width=0.495\textwidth,page=2]{figs-paper/np-corrections-jetpt.pdf}
  \caption{
Monte Carlo study of the impact of hadronisation and underlying event (UE) on the ungroomed $\pt$ distribution (on the left) and on the $\ptg$ distribution.}
  \label{fig:np-effect-inclusive}
\end{figure}

Moving back to $pp$ collisions, we study how the nature of the observable, IRC safety for $\pt$ and Sudakov safety for $\ptg$, correlates with the size of non-perturbative corrections due to the hadronisation process and to multiple parton interactions, i.e.\ the underlying event (UE).
Although the question of a field-theoretical understanding of non-perturbative corrections and their interplay with substructure algorithms is of great interest, in this study we limit ourselves to a phenomenological approach based on Monte Carlo parton-showers simulations. 
In order to minimise potential bias due to a particular non-perturbative model, we use a variety of parton showers with different tunes, namely 
the  AUET2~\cite{ATLAS:2011gmi} tune of \herwig~6.521~\cite{Corcella:2000bw,Corcella:2002jc},
the Z2~\cite{Field:2010bc} and Perugia~2011~\cite{Skands:2010ak,Cooper:2011gk} tunes of \pythia~6.428~\cite{Sjostrand:2006za}, the 4C~\cite{Corke:2010yf} and the Monash~13 tune~\cite{Skands:2014pea} of \pythia~8.223~\cite{Sjostrand:2007gs}.
The results of this study are presented in Fig.~\ref{fig:np-effect-inclusive}, where the plot on the left shows the ungroomed $\pt$ spectrum, while the one on the right the $\ptg$ distribution.  
In each plot, we show two sets of curves. The first set (labelled
``hadronisation'' on the plots) represents, for each Monte Carlo, the
ratio between hadron-level and parton-level results, without UE. The
second set (labelled ``UE'') instead shows the ratio of hadron-level
results with and without the UE contribution. The $\pt$ plot shows all
the features we would expect from an IRC safe
observable. Non-perturbative corrections are suppressed by negative
powers of the jet transverse momentum. Moreover, since we are dealing
with high-$p_t$ jet with a fairly large radius ($R=0.8$) hadronisation
corrections are rather small~\cite{Dasgupta:2007wa}. The Sudakov-safe
$\ptg$ distribution instead exhibits larger hadronisation corrections,
which do not appear to be power suppressed~\cite{Larkoski:2015lea}.
On the other hand, as perhaps expected in the presence of a groomer, we note that
$\ptg$ is less sensitive to the UE contribution than $\pt$, especially at
moderate transverse momentum. We can therefore expect that $\ptg$
will be more resilient against pile-up (not considered here), which
has a structure similar to the UE.
To cast more light on Sudakov-safe observables, it would be
interesting to investigate analytically the structure of hadronisation
corrections to the $\ptg$ cross-section, using e.g. techniques from
Ref.~\cite{Dasgupta:2007wa}.

In this study we are primarily interested in jet mass distributions,
while we only use the jet cross section for normalisation
purposes. Measuring a non-vanishing mMDT mass resolves a two-prong
structure within the jet, thus acting as an angular cut-off and
regulating the collinear divergence. This means that the unnormalised
distribution
\begin{align} \label{distr-ptg}
\frac{d \sigma}{ d \rho}(\rho;\zc,p_{t1},p_{t2})&= \int_{p_{t1}}^{p_{t2}} d \ptg \frac{d^2 \sigma}{d \ptg d \rho}, \quad \text{with} \quad \rho=\left(\frac{m}{\ptg R} \right)^2,
\end{align}
is IRC safe. However, as we shall see in the following section, the
resulting all-order structure is different compared to the one
previously described and rather cumbersome. We also note that, because
the difference between $\pt$ and $\ptg$ is $\mathcal{O}(\zc)$, if we
choose to use $\ptg$ we are forced to work at finite $\zc$.

As a final note, we point out that despite its issues related to IRC
safety, $\ptg$ shows some interesting properties in perturbative QCD.
For example, it is directly related to the ``energy loss''
distribution computed in Ref.~\cite{Larkoski:2014wba} in the small
$\zc$ limit. Modulo small corrections induced by the running of the
coupling, the energy loss distribution --- i.e. the $\ptg$ distribution
at fixed $\pt$ --- is independent of $\alpha_s$ and of the colour
factor of the parton initiating the jet. 
We discuss this briefly in the context of the $\pt$ jet cross-section
in Appendix~\ref{app:ptmmdt-perturbative}.

\subsection{Fixed-order structure of the mass distribution}\label{sec:fo-ptmmdt}

In order to better understand the structure of the mass distribution with $\ptg$ we analytically calculate Eq.~(\ref{distr-ptg}) to LO and NLO, in the collinear limit.
We start with a jet of momentum $\pt$.
 At $\mathcal{O}(\as)$ the jet is made of at most two partons. If one of them is groomed away by mMDT, then the resulting groomed jet is massless. Thus, in order to have a non-vanishing mass, the emission must pass the $\zc$ condition, leading to $\ptg=\pt$. Therefore, the LL distribution at LO is the same for the two transverse momentum choices and it reads (see also Ref.~\cite{Dasgupta:2013via})
 \begin{align}\label{lo-ptg}
\rho \frac{d \sigma^\text{LL,LO}}{ d \rho}(\rho;\zc,p_{t1},p_{t2})&= \int_{p_{t1}}^{p_{t2}} d \pt
 \left[\sigma_q(\pt) R'_q +\sigma_g(\pt) R'_g \right],
 \end{align}
 where $R'_{q(g)}$ have been defined in
 Section~\ref{sec:llresum}.

The situation changes when we move to NLO. 
We consider the sum of the double real emission diagrams and the real-virtual contributions, while the double virtual one only gives vanishing masses.
At NLO we have different colour structures. For convenience, we explicitly consider the $C_F^2$ contribution, which originates from the independent emission of two collinear gluons $1$ and $2$ off a quark leg. Analogous results can be obtained for the other colour structures.
Because we are interested in the LL contribution, we can order the two
emissions in angle, i.e. $\theta_1\gg \theta_2, \theta_{12}$.
The relevant contributions correspond to the situation where gluon 2
is real (and dominates the mMDT jet mass) and the large-angle gluon 1
is either real and groomed away, or virtual. 
The only difference with respect to our calculation in the $\pt$ case
(and of Ref.~\cite{Dasgupta:2013via}) is that here we further have to
make sure that the measured $\ptg$ falls in the transverse momentum
bin under consideration, say
${p_{t1}}<\ptg<{p_{t2}}$. Assuming for the moment that
${p_{t1}}<\pt<{p_{t2}}$, we therefore have the
additional constraint on the double-real emission contribution that
$\ptg=(1-z_1) \pt$ still falls in the same transverse momentum bin. We
thus have
\begin{align}\label{nlo-ptg-start}
\rho \frac{d \sigma^{\text{LL,NLO},C_F^2a}}{ d \rho}&= \left(\frac{\as C_F}{\pi}\right)^2 \rho \int_{p_{t1}}^{p_{t2}} d \pt \,
 \sigma_q(\pt)  \\ &\cdot \int_0^1 \frac{d \theta_1^2}{\theta_1^2} \int_0^1 d z_1 \, p_{gq}(z_1) 
 \Big[\Theta\left(\zc>z_1 \right) \Theta\left((1-z_1) \pt > p_{t1} \right)-1 \Big] \nonumber \\ &\cdot
 \int_0^1 \frac{d \theta_2^2}{\theta_2^2} \int_0^1 d z_2 \, p_{gq}(z_2) \Theta\left(z_2 >\zc \right)
  \Theta\left(1- z_2 >\zc \right) \Theta\left( \theta_1^2> \theta_2^2\right)\delta\left( \rho - z_2 \theta_2^2\right).\nonumber
\end{align}
After some algebra, the distribution in the $\rho<\zc$ region can be
written in terms of the $R_i$ functions previously defined
\begin{align}\label{nlo-ptg-cntd}
\rho \frac{d \sigma^{\text{LL,NLO},C_F^2a}}{ d \rho}&= \int_{p_{t1}}^{p_{t2}} d \pt \,
 \sigma_q(\pt) R'_q \Big[-R_q-R_{q\to g} \Big] \\ &-
 \int_{p_{t1}}^{\min\left[p_{t2},\frac{p_{t1}}{1-\zc} \right]}d \pt \,
 \sigma_q(\pt) \;  R'_q \;
  \frac{\as C_F}{\pi} \log\frac{1}{\rho}\int_{1-\frac{p_{t1}}{\pt}}^{\zc} d z_1 \, p_{gq}(z_1). \nonumber
\end{align}
We note that the first contribution coincides with the expansion of
the resummation formula Eq.~(\ref{eq:matrix}) to second
order. However, the second term, proportional to
$\alpha_s^2\log(1/\rho)$, is a new LL contribution that signals the
different all-order structure of the mass distribution with
$\ptg$. Note that we have put a label $a$ in
Eqs.~(\ref{nlo-ptg-start}) and~(\ref{nlo-ptg-cntd}) because there is
actually a second configuration that contributes, namely when the
ungroomed jet has $\pt> p_{t2}$. If the first emission
is groomed away, we may end up with $\ptg< p_{t2}$, so that
this contribution has now leaked into the lower bin. For a
quark-initiated jet with two gluon emissions, this results
into an additional LL piece:
\begin{align}\label{nlo-ptg-b}
\rho\frac{d \sigma^{\text{LL,NLO},C_F^2b}}{ d \rho}&= \int_{p_{t2}}^\frac{p_{t2}}{1-\zc} d \pt \,
 \sigma_q(\pt) \; R'_q \;  \frac{\as C_F}{\pi} \log\frac{1}{\rho}\int_{1-\frac{p_{t2}}{\pt}}^{\zc} d z_1 \, p_{gq}(z_1).
\end{align}

\subsection{Resummation} \label{sec:resum-ptmmdt}
In order to resum the groomed jet mass spectrum in the case of the $\ptg$ selection we have to generalise the calculation described in the previous section to all orders.  Clearly, the situation is much more complicated than the $\pt$ case chiefly because the value of $\ptg$ is determined by all the emissions that fail the mMDT condition and therefore our calculation must keep track of them.  
Because of this complication we are not able to find simple analytic
expressions that capture the all-order behaviour, nevertheless we can
achieve LL accuracy in the groomed mass distribution using an approach
based on generating functionals~\cite{Ellis:1991qj,Dokshitzer:1991wu}
and, in particular, the application of this formalism to the
description of the angular evolution of jets with small
radius~\cite{Dasgupta:2014yra,Dasgupta:2016bnd}.

We start by defining an evolution variable which is closely related to
the angular scale $\theta$ at which we resolve a jet
\begin{align}\label{t-def}
t= \int_{\theta^2}^{1} \frac{d {\theta'}^2}{{\theta'}^2} \frac{\as({\theta'}\pt R)}{2\pi}
 = \frac{1}{2\pi\beta_0}\,\log\bigg(\frac{1}{1+2\as\beta_0\log(\theta)}\bigg)
 = \frac{\as}{2\pi}\,\log\bigg(\frac{1}{\theta^2}\bigg) + {\cal{O}}(\as^2),
\end{align}
with, as before, $\as=\as(p_tR)$.
This definition of $t$ includes leading collinear logarithms induced
by the running of the QCD coupling when going to small angles.
When mMDT (and more generically Soft Drop) recurses to smaller and
smaller angular scales, the corresponding value of evolution variable
$t$ increases until it reaches a non-perturbative value
$t_\text{max}$.
Thus, by considering successive $1\to2$ angular-ordered splittings, we can write down LL evolution equations for a generating functional associated to a quark, $Q(x,t)$, or to a gluon $G(x,t)$, where $x$ is the momentum fraction.
The relevant evolution equations were derived in
Ref.~\cite{Dasgupta:2014yra}. The only difference here is that after
each splitting we follow the branch with the highest transverse
momentum, as it is appropriate for the mMDT algorithm. We obtain
\begin{subequations} \label{gen_func}
\begin{align}
\frac{d}{d t}Q(x,t)&= 2C_F \int_0^1 dz\, p_{gq}(z) \left [Q\left( (1-z) x \right)\Theta\left(z< \frac{1}{2}\right)+
G\left( z x \right)\Theta\left( z>\frac{1}{2}\right) - Q(x,t)\right], \\
\frac{d}{d t}G(x,t)&= 2C_A \int_0^1 dz\, 
\bigg[ \frac{1}{2} p_{gg}(z)  G\big(\text{max}(z,1-z) x,t\big) +\frac{T_R
                     n_f}{C_A}p_{qg}(z)  Q\big(\text{max}(z,1-z)
                     x,t\big) \nonumber \\
 & \phantom{= 2C_A \int_0^1 dz\, [}     - p_{x g} (z) G(x,t) \bigg].
\end{align}
\end{subequations}
These equations can be implemented numerically under the form of a
Monte Carlo generator producing angular-ordered (from large angles to
small ones) parton branchings. Compared to the implementation used
in~\cite{Dasgupta:2014yra}, the only difference is that the successive
branchings follow the hardest of the two partons obtained at the
previous step of the showering. We record the angle $\theta$ and
momentum fraction $z$ of all the emissions.

In order to obtain the mMDT mass spectrum, two extra ingredients are
needed: firstly, we need to impose the mMDT condition and, secondly, we
should impose an ordering in invariant mass rather than an ordering in
angle.
Since mMDT proceeds by declustering a C/A tree, imposing the mMDT condition on our angular-ordered events is
trivial: we simply search for the first emission that satisfies
$\zc<z<1-\zc$. From the momentum fractions of all the previous
emissions, i.e. those at larger angles, we can then reconstruct the
momentum fraction groomed away by the mMDT procedure and thus $\ptg$.
Then, once we have reached an emission that passes the mMDT condition,
we investigate all the emissions to find the one that dominates the
mass. If these emissions have angles $\theta_i$, obtained by
inverting Eq.~(\ref{t-def}), and momentum fractions $z_i$, we take,
to LL accuracy, $\rho = \text{max}_i[\text{min}(z_i,1-z_i)\theta_i^2]$.
In particular, it is worth pointing out that we can use the momentum
fraction $z_i$, relative to each branching, instead of the actual
momentum of each parton with respect to the initial jet. This is simply because
the difference between the two does not generate any logarithmic
enhancement.\footnote{Similarly, we can wonder why, once we have an
  emission satisfying the mMDT condition and the de-clustering
  procedure stops, we keep generating branchings only on the hardest
  branch.
  This is simply because further branchings on a soft branch would
  never dominate the jet mass and can therefore be neglected.
  This would not be valid for observables sensitive to secondary
  emissions, like $N$-subjettiness with $N>1$, for which all
  branchings should be included at angles smaller than the first
  branching which passes the mMDT condition.}
Finally, since the resummation is obtained from a Mote Carlo event
generator, it can directly be interfaced with \nlo at Born-level to
obtain predictions for the jet mass cross-section.

Before we present matched results, we note that, compared to the
resummation done in the previous section for the $\pt$ case, the use
of Eq.~(\ref{t-def}) implies  that we are neglecting a factor
$z$ in our choice of the scale of the running coupling.
This means that we are not including running-coupling effects in the
double-logarithmic small-$\zc$ contributions. 
This approximation can be explicitly studied in the context of a
selection on $\ptg$ and we show in Appendix~\ref{app:zinalphas} that
this only have a modest impact on the final results.

\subsection{Matching and perturbative results} \label{sec:ptmmd_matching}

 \begin{figure}[t]
  \centering
  \includegraphics[width=0.495\textwidth,page=1]{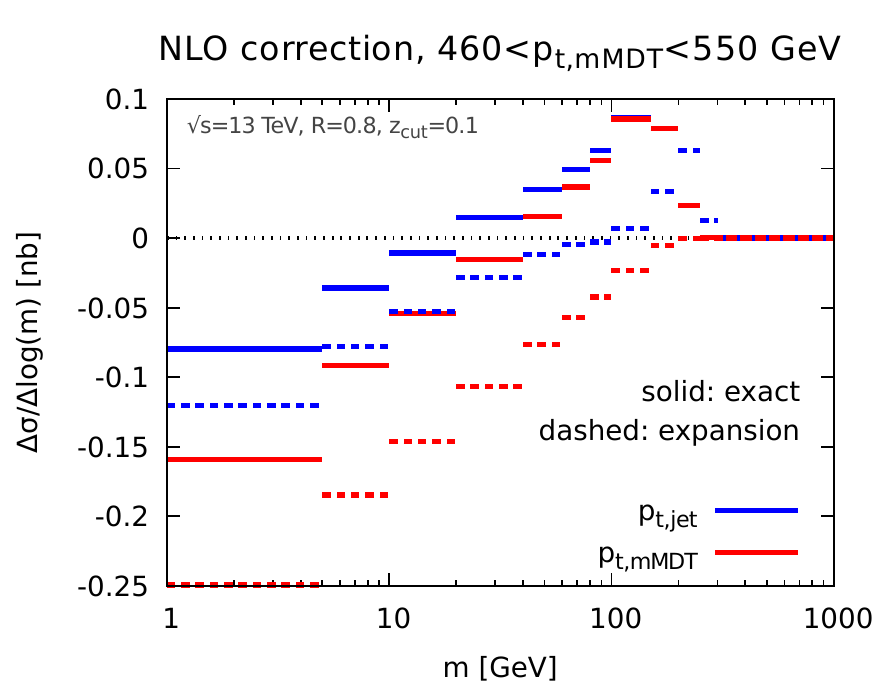}
  \includegraphics[width=0.495\textwidth,page=2]{figs-paper/ptmmdt-expansion-both.pdf}
  \caption{Comparison between the full NLO correction (solid) to the jet mass distribution and the $\ord(\as^2)$ expansion of the LL resummation (dashed) for both $\pt$ (blue) and $\ptg$ (red) in two different transverse momentum bins.}
  \label{fig:groomed-pt-fo-checks}
\end{figure}

As for the case of the ungroomed $\pt$, an accurate description valid
both in the $\rho \ll 1$ region and in the $\rho \sim 1$ region
requires the matching of our LL resummation to a fixed-order
calculation. As before, the latter is obtained using \nlo. We note that at LO, the
results are identical to the ones obtained in the $\pt$ case,
Section~\ref{ptjet:results}.

In order to match fixed-order and resummed calculations we have to work out the expansion of the resummed
cross-sections to LO and NLO. This can be obtained by expanding
Eq.~(\ref{gen_func}) to first and second order in $\as$.
In practice, we have found more convenient to reuse here the same code as in
Ref.~\cite{Dasgupta:2014yra}, with minor modifications to
  include additional information about the successive branching angles
  and momentum fractions as well as simplifications related to the
  fact that we do not have to include splittings in the soft branch.
For fixed $p_t$, we have checked our numerical results against an
explicit analytic calculation.
Note that at NLO, i.e. at ${\cal O}(\as^2)$, we should include both a
contribution coming from two emissions (see also the earlier
discussion in Section~\ref{sec:fo-ptmmdt}) as well as a
running-coupling correction coming from the expansion of Eq.~(\ref{t-def})
to ${\cal O}(\as^2)$.

We compare the expansion of the LL resummation to $\ord(\as^2)$
against the exact \nlo NLO correction in
Fig.~\ref{fig:groomed-pt-fo-checks}, for both $\pt$ (blue) and $\ptg$
(red) and for two different transverse momentum bins.
We first note that at small mass the expansion of the resummed
distribution has the same slope of the corresponding fixed-order,
meaning that we do indeed control the ${\cal O}(\as^2\log(1/\rho))$
contribution, as expected from our LL calculation.
More interestingly, Fig.~\ref{fig:groomed-pt-fo-checks} shows
explicitly that the mass distributions obtained in the $\ptg$ and
$\pt$ cases have different slopes at small mass, i.e. different
${\cal O}(\as^2\log(1/\rho))$ contributions. This means that the two
distributions already differ at the LL level.
The difference in slope is captured by our analytic calculation and is
due to the effects already discussed in Section~\ref{sec:fo-ptmmdt}.

\begin{figure}[t]
  \centering
  \includegraphics[width=0.495\textwidth,page=1]{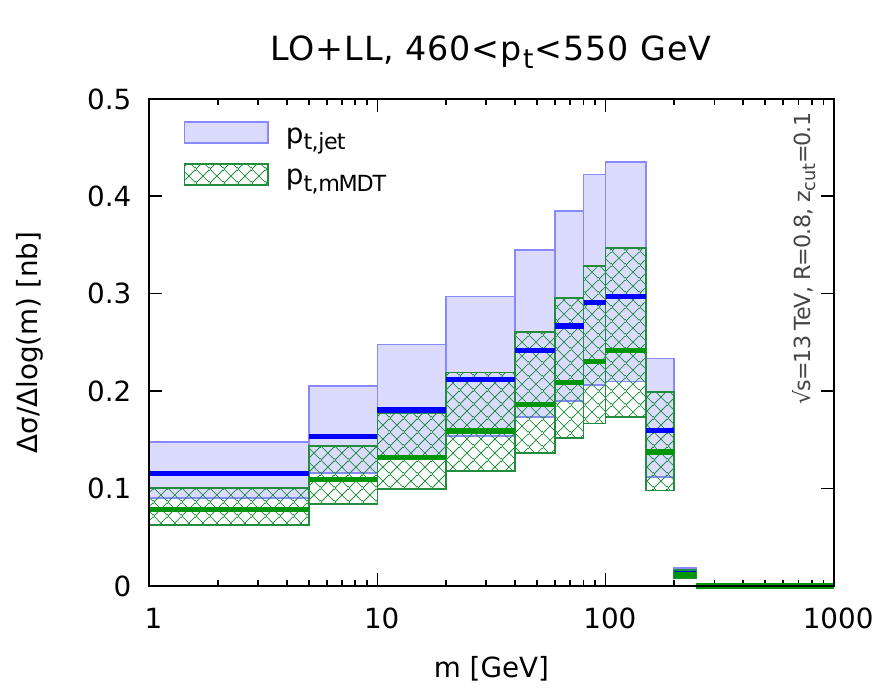}
  \includegraphics[width=0.495\textwidth,page=2]{figs-paper/ptmmdt-v-ptjet-LO.pdf}
  \includegraphics[width=0.495\textwidth,page=1]{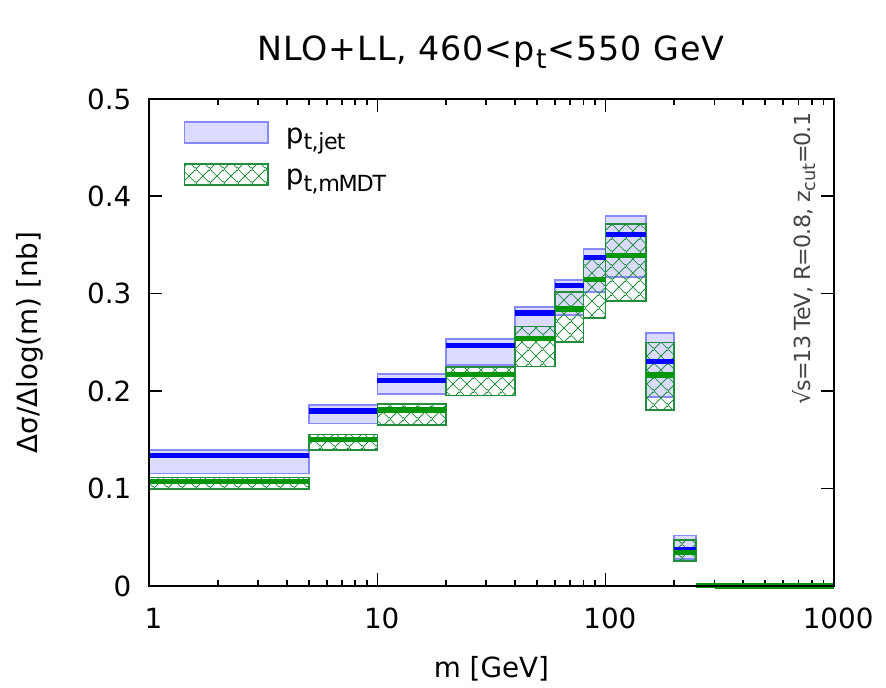}
  \includegraphics[width=0.495\textwidth,page=2]{figs-paper/ptmmdt-v-ptjet-NLO.pdf}
  \caption{
  In this figure we compare resummed and matched jet mass distributions in the case of ungroomed $\pt$ selection (blue) or groomed $\ptg$ selection (green).}
  \label{fig:distr-groomed}
\end{figure}

We are now ready to discuss the matching itself. We adopt the
multiplicative matching scheme introduced in Eq.~(\ref{match1}). 
Our results are shown in Fig.~\ref{fig:distr-groomed} for the
(unnormalised) jet-mass cross-section. The hatched (green) curves are
the results obtained for the $\ptg$ case and they are compared to the
results already obtained in Section~\ref{ptjet:results} shown in
shaded blue.
The plots on the top are for LO+LL, while the ones at the bottom for
NLO+LL. We pick the same representative bins in transverse momentum as
before, namely $460< p_t< 550$~GeV and $p_t>1300$~GeV, with $p_t$
being either $\ptg$ or $\pt$.
Fig.~\ref{fig:groomed-pt-fo-checks}.
The cross-sections are smaller for the $\ptg$ case than for the $\pt$
case, mostly because the overall inclusive jet cross-section is
smaller. This is related to the loss of transverse momentum when
applying the mMDT procedure, which is also discussed in
Appendix~\ref{app:ptmmdt-perturbative}.
We also see, in particular in the NLO+LL results for the high-$p_t$
bin, that $\ptg$ distributions decrease slightly faster than the $\pt$
ones, at small mass. This feature was already observed in
Fig.~\ref{fig:groomed-pt-fo-checks} and it can be attributed to the
presence of extra logarithmic contributions, which further suppress
the distribution at small values of the mass.

We note that due to the IRC unsafety of the $\ptg$ jet cross-section,
the normalisation of the fixed-order jet mass distribution is ill-defined. 
The resummed and matched cross-sections could simply be
normalised to unity but we found that this procedure tends to clearly
underestimate the size of the perturbative uncertainty and is
potentially dangerous as it relies on the computation of the resummed
cross-section down to very small masses where non-perturbative effects
are dominant.
We have therefore decided to present only predictions for the unnormalised
distributions.

\section{Non-perturbative corrections}\label{sec:np-corrections}

\begin{figure}[t]
  \centering
  \includegraphics[width=0.495\textwidth,page=1]{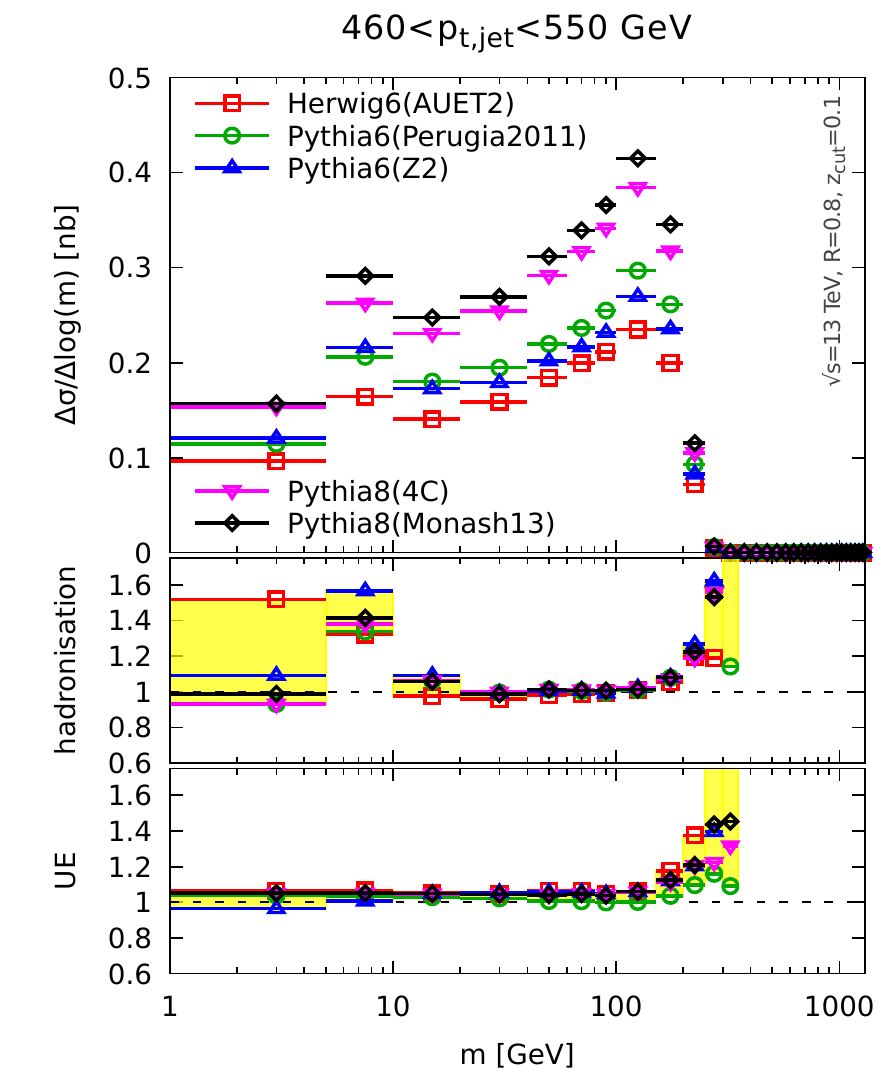}
  \includegraphics[width=0.495\textwidth,page=2]{figs-paper/mass-distrib-np.pdf}
  \caption{ The top plots show the groomed jet mass distribution for
    $460< p_t< 550$~GeV, with hadronisation and the underlying event,
    for different Monte Carlo parton showers. The plot on the left is
    for the ungroomed $\pt$, while the one of the right for $\ptg$.
    The bottom plots show the ratios hadron-to-parton level and
    with-to-without the underlying event.  }
  \label{fig:np-effect-mass}
\end{figure}

\begin{figure}
  \centering
  \includegraphics[width=0.495\textwidth,page=3]{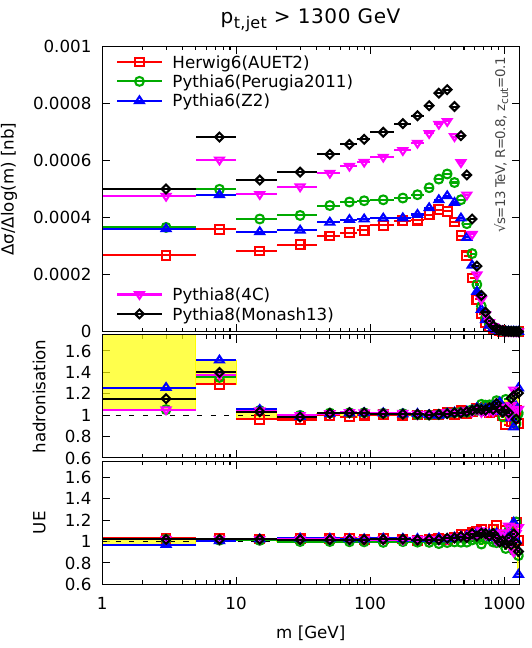}
  \includegraphics[width=0.495\textwidth,page=4]{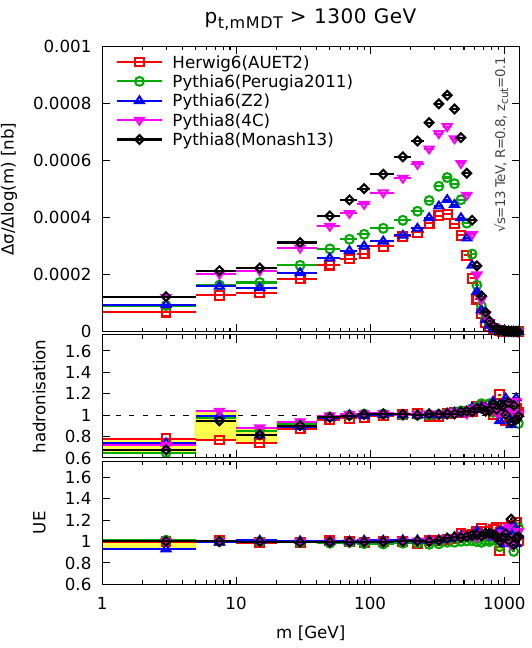}
  \caption{
Same as Fig.~\ref{fig:np-effect-mass} but for the bin $p_t>1300$~GeV.}  \label{fig:np-effect-mass-2}
\end{figure}

In this section we perform a Monte Carlo study of non-perturbative
contributions considering effects coming from the hadronisation
process as well as from the underlying event.
In order to study non-perturbative corrections to the jet mass
distribution we consider the same set of Monte Carlo tunes used for
studying the $p_t$ spectra in Section~\ref{sec:ptgroomed}. As usual,
we consider two representative transverse momentum bins.
In Fig.~\ref{fig:np-effect-mass} we consider $460< p_t< 550$~GeV,
while in Fig.~\ref{fig:np-effect-mass-2} we consider
$p_t>1300$~GeV. In both cases, the plots on the left refer to the
ungroomed $\pt$ selection, while the ones on the right refer to the
$\ptg$ case.

In the top plots we show the (unnormalised) jet mass distributions as
obtained from each Monte Carlo program. The striking feature is the
huge discrepancy between these results, even at large masses. In
particular, the predictions obtained with the most recent \pythia~8
tunes appear to be a factor of 2 larger than the other tunes in the
region of interest for this study.
This performance of standard parton shower tools, worrisome at first
glance, should be put in parallel with our LO+LL results (see
e.g. Fig.~\ref{fig:distr-groomed}) which exhibit a similar uncertainty
band.
This indicates the need to match the parton shower with NLO
fixed-order matrix elements.

In the bottom plots of Figs.~\ref{fig:np-effect-mass}
and~\ref{fig:np-effect-mass-2} we instead show, for each Monte Carlo,
the ratio of hadron-to-parton level results (labelled
``hadronisation") and the ratio with-to-without the underlying event
contribution (labelled ``UE").
We first note that in both the $\pt$ and $\ptg$ selection cases, the
groomed mass distribution has very small sensitivity to the underlying
event, as we expect from mMDT being an (aggressive) groomer. 
This contribution becomes more sizeable at large masses essentially
because the effective jet radius becomes larger.  Moreover, this
effect is more visible in the moderate $p_t$ bin since the
power-suppression in the hard scale of the process becomes weaker.
Hadronisation corrections have instead a different shape for the $\pt$
and $\ptg$ selections, most likely stemming from the different
properties of the underlying transverse momentum distribution. For the
$\pt$ case, hadronisation corrections are sizeable in the low mass
bins, with a peculiar peak in the 5-10 GeV bin, and at very large
masses, close to the end-point region.
For both small and large masses, this also comes with a larger spread
of the hadronisation corrections across the generators and tunes.
However, there exists a rather large region in mass, increasing in
size as $\pt$ grows, where these contributions are genuinely
small.
Hadronisation corrections have a similar size in the $\ptg$ selection
case. This is not unexpected because the mass distribution is itself
IRC safe.  However, they do exhibit a rather different shape.
They come with opposite sign at small masses and
appear to be non-negligible in a wider region of the mass
distribution, similarly to what was already noticed in
Section~\ref{sec:ptmmdt-safety} for the jet cross-section.

Given the large kinematic range over which the non-perturbative
corrections appear to be small, upcoming LHC data could bring valuable
constraints on the perturbative aspects of parton showers.
Additionally, the behaviour at low mass, with very little sensitivity to the
underlying event, could help constraining hadronisation models. For
example, measurements on both quark and gluon-enriched jet samples
would be complementary to the quark-dominated LEP data currently
used to tune hadronisation models~\cite{Badger:2016bpw,QuarkGluon}.

In practice, for this study, we use the above Monte Carlo results to estimate the size
and the uncertainty of non-perturbative corrections on the groomed
mass distribution.
For each Monte Carlo generator and tune we construct the ratio
particle-level, i.e.\ hadronisation with UE, to parton-level, in each
mass and transverse momentum bin.
We take the average value of this ratio as a correction factor to
apply to the perturbative NLO+LL results obtained in the previous
sections.  We take the envelope of the corrections across different
generators and tunes as an estimate of the non-perturbative
uncertainty, which we add in quadrature to the perturbative
uncertainty. We consider this solution an acceptable and rather
conservative estimate of non-perturbative contributions, lacking a
detailed, field-theoretical study of these corrections in the presence
of grooming algorithms.

\section{Final results}\label{sec:final}

We can now present our final results for the groomed jet mass
distribution for both the $\pt$ and $\ptg$ selection. Our perturbative
results, which are accurate to NLO+LL, are multiplied by a bin-by-bin
(in both mass and transverse momentum) non-perturbative correction
factor obtained from Monte Carlo parton showers.
The total uncertainty is taken as the sum in quadrature of the
perturbative and non-perturbative uncertainties.
The former is obtained by varying renormalisation, factorisation, and
resummation scales as described in Section~\ref{sec:ptjet} and taking
the envelope of the result; the latter by considering the envelope of
the five different Monte Carlo generators and tunes.

\begin{figure}
  \centering
  \includegraphics[width=0.495\textwidth,page=1]{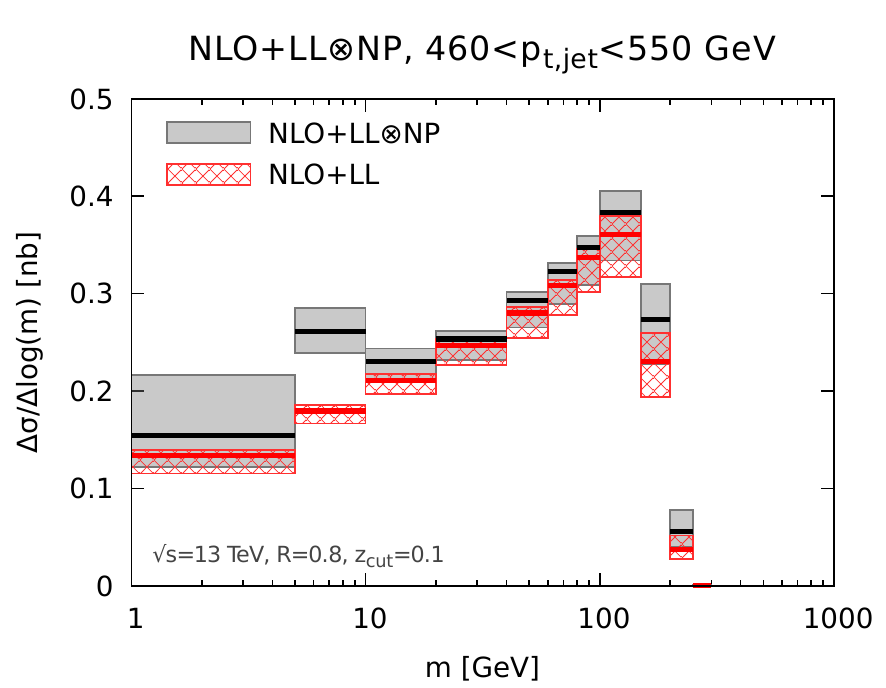}
  \includegraphics[width=0.495\textwidth,page=2]{figs-paper/ptjet-matched-np.pdf}
  \caption{ Final results for the jet mass distribution in the case of
    the ungroomed $\pt$ selection. The perturbative calculation is
    performed at NLO+LL and non-perturbative corrections are included
    as a multiplicative factor obtained from Monte Carlo parton
    showers. Perturbative uncertainties are obtained varying
    renormalisation, factorisation and resummation scales as detailed
    in Section~\ref{sec:ptjet}. Non-perturbative uncertainties are
    obtained considering the spread of five different Monte Carlo
    tunes, as detailed in
    Section~\ref{sec:np-corrections}. Perturbative and
    non-perturbative uncertainties are added in quadrature.}
  \label{fig:final}
\end{figure}

\begin{figure}
  \centering
  \includegraphics[width=0.495\textwidth,page=1]{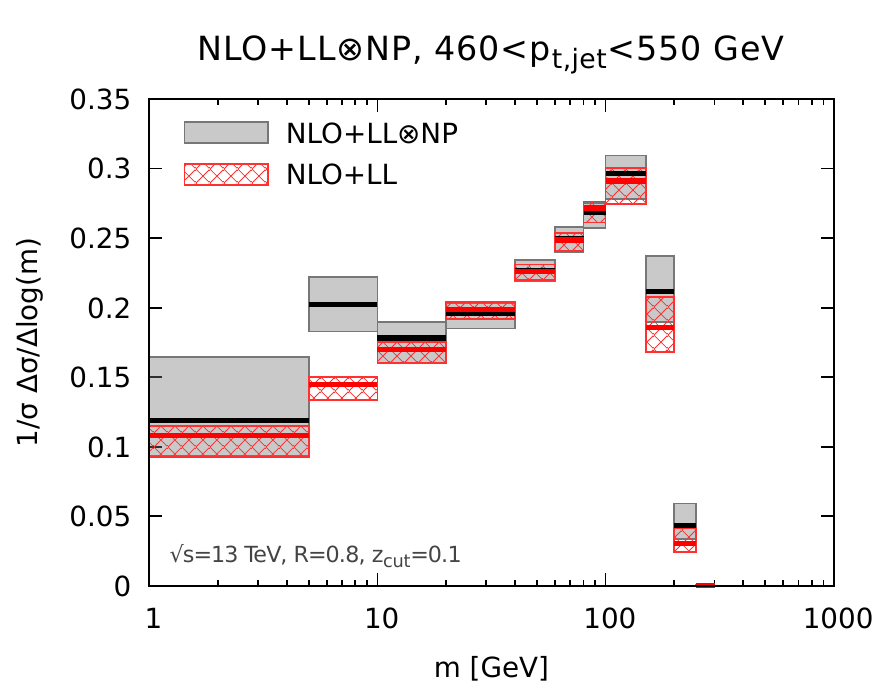}
  \includegraphics[width=0.495\textwidth,page=2]{figs-paper/ptjet-matched-np-norm.pdf}
  \caption{
Final results at NLO+LL, with non-perturbative corrections, for the normalised jet mass distribution, in the case of the ungroomed $\pt$ selection.}  \label{fig:final-norm}
\end{figure}

\begin{figure}
  \centering
  \includegraphics[width=0.495\textwidth,page=1]{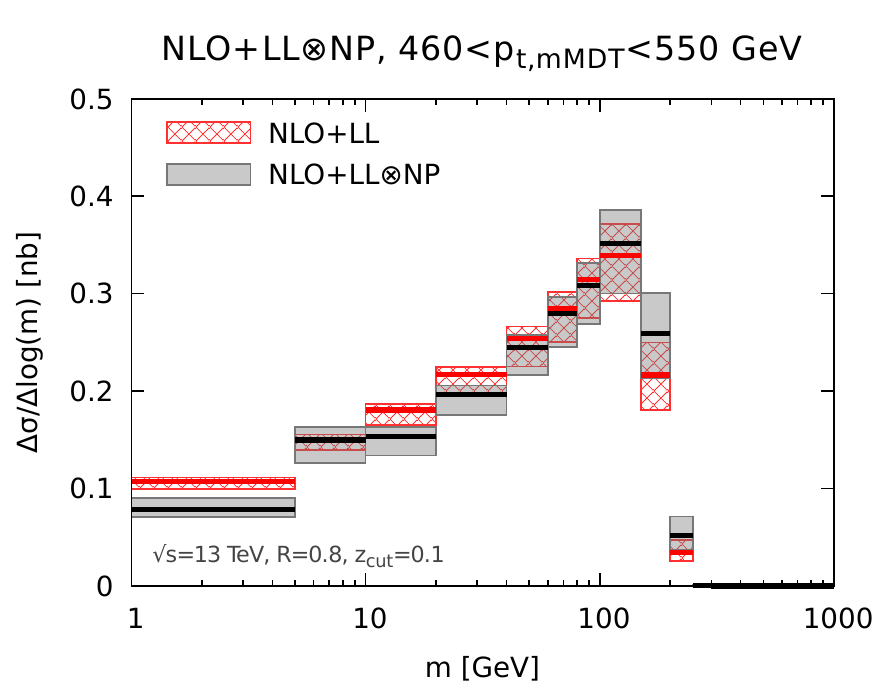}
  \includegraphics[width=0.495\textwidth,page=2]{figs-paper/ptmmdt-matched-np.pdf}
  \caption{
Final results at NLO+LL, with non-perturbative corrections, for the $\ptg$ selection.}  \label{fig:final-2}
\end{figure}

Fig.~\ref{fig:final} and Fig.~\ref{fig:final-norm} show the results (in black, with grey uncertainty bands)
for the ungroomed $\pt$ selection in the two representative transverse
momentum bins: $460< \pt< 550$~GeV and $\pt>1300$~GeV. The former is
the jet mass distribution, while the latter is normalised to the NLO
jet cross-section in the appropriate transverse momentum bin.
Similarly, in Fig.~\ref{fig:final-2} we show our final results for the
$\ptg$ selection. As discussed in the paper, the NLO jet cross section
is not well-defined in this case, so we only present unnormalised
distributions.
For comparison, we also show in red the purely perturbative NLO+LL
results with their uncertainties.
As previously noted, non-perturbative corrections are sizeable (with
large uncertainties) in the first few mass bins ($m\lesssim10$~GeV)
and at very large masses, close to the end-point region.
Nevertheless, there exists a region in mass, which increases in size
as $\pt$ grows, where non-perturbative effects are genuinely small and
a meaningful comparisons between experiments and perturbation theory
can be performed. However, we have found that, when we consider
normalised distributions in Fig.~\ref{fig:final-norm}, the uncertainty
related to these non-perturbative contributions is, at best, of the
same order as the NLO+LL perturbative calculation.

The above results clearly demonstrate the value of jet substructure
algorithms to perform phenomenological studies in QCD. In particular,
the region in mass where non-perturbative contributions are genuinely
small offers an opportunity to test the modeling of perturbative
radiation in analytic resummations and parton showers. In that
respect, one could even consider the possibility to use experimental
data in this mass region for a novel measurement of the strong
coupling. On the other hand, the lower mass bins, which are sensitive
to hadronisation but have small UE contaminations, can be used to test
(and tune) the hadronisation models of Monte Carlo event
generators. To this purpose, it will be also interesting to extend
this analysis to different jet shapes and angularities, and to
different processes, e.g. \ Z+jet, which have different sensitivity to
QCD radiation, both at the perturbative and non-perturbative level.
Furthermore, while the $\pt$ selection is under better theoretical
control and should be amenable to a higher logarithmic accuracy, we
think that the $\ptg$ case also offers many interesting physics
opportunities. While the jet mass distribution is IRC safe in both
cases, the underlying $\ptg$ itself is not. Detailed studies of these
types of observables will improve our understanding of Sudakov safety. 
Furthermore, the two
transverse-momentum selections exhibit different sensitivities to
non-perturbative effects (especially hadronisation). Such a
measurement could therefore shed some light on power corrections for
Sudakov-safe observables and be of further help for Monte Carlo
tuning.
%

\section{Conclusions}\label{sec:conclusion}
 
In this paper we have considered the production of hadronic jets in
proton-proton collisions and studied the invariant mass distributions
of groomed jets, focusing on the mMDT algorithm, sometimes also
referred to as Soft Drop with the angular exponent set to zero.
Our calculation is double-differential in jet mass and transverse
momentum and fully takes into account the kinematic cuts of an
upcoming CMS measurement at $\sqrt{s}=13$~TeV. We present our results
as jet mass distributions in different transverse-momentum bins.

Jet mass distributions receive logarithmic corrections originating
from the emissions of soft and/or collinear partons. However, the
presence of a grooming algorithm mitigates the contributions from the
soft region of phase-space.
The resulting mMDT mass distribution is single-logarithmic with the
logarithmic enhancements only stemming from the hard-collinear region.
We have resummed this contribution to LL accuracy.
In doing so we have lifted the small-$\zc$ approximation which has
been used in other studies aimed at a higher logarithmic
accuracy~\cite{Frye:2016okc,Frye:2016aiz}.
In order to also describe the high-mass tail of the distribution we
match to fixed-order matrix elements at NLO using the program \nlo.

We have considered two different choices for the transverse momentum
selection. The first option consists in selecting and binning the jets
according to their transverse momentum before grooming, namely $\pt$,
while in the second one the transverse momentum after grooming $\ptg$
is used. We note that a calculation performed in the small-$\zc$ limit
cannot resolve this difference, as the two are equal at $\zcut=0$.

We have found that the $\pt$ selection is better suited for theoretical
calculations and the resulting resummation has a relatively simple
form that can be, in principle, extended to higher-logarithmic
accuracy. Moreover, for the typical choice $\zc=0.1$, finite $\zc$
corrections, although formally entering already at LL accuracy,
appear to be very small.
This justifies the small-$\zc$ approximation used in
Refs.~\cite{Frye:2016okc,Frye:2016aiz} to achieve higher logarithmic
accuracy.
However, the finite $\zcut$ corrections would inevitably
increase for larger values of $\zcut$. Also, it would be interesting
to achieve a complete picture at NLL accuracy, including the finite
$\zcut$ corrections, even though our findings in this paper suggest
that the latter would be small.
We have also found that logarithms of $\zc$ give a non-negligible
contribution, thus indicating the necessity of their resummation.
We have also studied the perturbative uncertainty of our calculation,
observing that matching to NLO greatly reduces the theoretical
uncertainty especially in the case of unnormalised distributions.
Finally, we have studied non-perturbative contributions from
hadronisation and the underlying event using different Monte Carlo
parton showers.  Non-perturbative effects are reduced compared to the
ungroomed jet mass and only remain sizeable at low mass, where
hadronisation dominates, or at very large masses, close to the
end-point of the distribution.

The $\ptg$ selection has instead more theoretical issues but it can
also present some advantages from a phenomenological viewpoint. The
main theoretical complication stems from the fact that the $\ptg$ jet
spectrum is not IRC safe, but only Sudakov safe. The jet mass
distribution is itself safe, with the mass acting as a regulator for
collinear emissions, but the inclusive $\ptg$ cross-section is only
Sudakov safe.
Due to the complicated flavour structure of the all-order resummation,
we were only able to arrive at a numerical resummation of the LL
contributions. A possible extension of our results to a higher
logarithmic accuracy is therefore expected to be difficult, even in
principle.
From a phenomenological
  viewpoint, it would be interesting to see whether the slightly smaller
  sensitivity to the underlying event of the $\ptg$ choice implies a
  smaller sensitivity to pileup. 

To summarise, in this work we have derived theoretical predictions for
the invariant mass distribution of jets groomed with mMDT, including a
study of the perturbative and non-perturbative theoretical
uncertainties.
The situation where distributions are computed in bins of the initial
(ungroomed) jet $p_t$ exhibit a simpler analytic structure, compared
to the case where the binning is done using the groomed jet $p_t$.
This means that the former is more likely to be amenable to a
theoretical calculation with higher logarithmic accuracy.
We look forward to comparing our calculations to
upcoming LHC measurements and extend our predictions to additional
observables.

\begin{acknowledgments}
We thank Mrinal Dasgupta, Andrew Larkoski, Sal Rappoccio, Gavin
Salam and Jesse Thaler for many useful discussions.
SM would like to thank IPhT Saclay for hospitality during the course
of this project.
The work of SM is supported by the U.S.\ National Science
Foundation, under grant PHY-1619867, All-Order Precision for LHC
Phenomenology.
GS's work is supported in part by the French Agence Nationale de la
Recherche, under grant ANR-15-CE31-0016 and by the ERC Advanced Grant
Higgs@LHC (No.\ 321133).
\end{acknowledgments}

\appendix 

\section{Details of the analytic calculation}\label{app:calc-all}

In this Appendix we give more detail about the calculations of the
resummation functions $R_i$ introduced in Section~\ref{sec:llresum}.

\subsection{Resummed exponents}\label{app:calc}

The splitting functions introduced in
Eqs.~(\ref{eqn:sudakov-integral-representation}) are defined as
\begin{eqnarray}
p_{gq} &=& \dfrac{1 + (1-z)^2}{2 z},\\
p_{qg}  &=& \dfrac{1}{2} \big(z^2 +(1-z)^2\big),\\
p_{gg} &=& \dfrac{2(1-z)}{z} + z(1-z),
\end{eqnarray}
and we have also defined the following combination
\begin{equation}
p_{xg} \equiv \dfrac{1}{2} p_{gg} + \dfrac{T_R n_f}{C_A} p_{qg}.
\end{equation}
The running coupling used in
Eqs.~(\ref{eqn:sudakov-integral-representation}) is computed at the
one-loop accuracy, namely
\begin{equation}\label{eq:coupling-1loop}
\as(\kappa)=\frac{\as(Q)}{1+2 \as(Q) \beta_0 \log \frac{\kappa}{Q}} \ .
\end{equation}
Our results are expressed in terms of $\as=\as(R\, \pt)$, evolved from
$\as(m_Z)=0.118$ with a two-loop approximation ($n_f=5$).\footnote{Our
  use of the two-loop running coupling to compute $\alpha_s$ at the
  hard scale comes from the fact that we ultimately match our
  resummed calculation to a NLO fixed-order calculation which itself
  uses the two-loop running coupling as obtained from the NLO CT14 PDF set~\cite{Dulat:2015mca}.}
Note that for the minimal jet mass of 1 GeV that we consider in
this paper and the variations of the renormalisation and resummation
scales, $\mu_R$ and $\mu_Q$, our perturbative results always remain
above the Landau pole.
We could decide to freeze the coupling at a scale $\mu_{\text{NP}}$
that we would vary around 1~GeV, and hence obtain an uncertainty associated 
to using perturbative QCD in a region sensitive to non-perturbative
effects. However this effect should be included already in our
estimate of the non-perturbative effects via the Monte-Carlo
simulations discussed in Section~\ref{sec:np-corrections}.

To obtain the results presented in the main text, we have written the
splitting functions entering the flavour-diagonal contributions as a
sum of two different contributions:
\begin{subequations} \label{eqn:splitting-split}
\begin{align}
  p_{gq}(z)
    & = \frac{C_F}{z}\Theta(z<e^{B_q}) + p_{gq}^{\text{(finite)}}(z),\\
  p_{xg}(z)
    & = \frac{C_A}{z}\Theta(z<e^{B_g}) + p_{xg}^{\text{(finite)}}(z).
\end{align} 
\end{subequations}
The cut-off at $z=e^{B_i}$ is such that the leftover finite part only
generates power corrections in $\zc$ while the $\log(1/\zc)$ and
constant terms are included in the first terms proportional to $1/z$.
Note that this will naturally produce distributions with an end-point
at $\rho=e^{B_i}$.
That said, the contribution from the first term can be integrated
straightforwardly and gives the ${\mathcal{R}}_i$ function given in
Eq.~(\ref{eq:sudakov}).

Next, we consider the contributions coming from the second term in
Eq.~(\ref{eqn:splitting-split}), as well as from the flavour-changing
contributions, which will be power-suppressed in $\zc$. For these, we
can safely ignore the factor $z$ in both the argument of $\alpha_s$
and the constraint $\Theta(z\theta^2>\rho)$. The $z$ and $\theta^2$
integration then factorise to give
\begin{equation}\label{eq:result-finite}
\text{finite part: }\int_\rho^{\zc} \frac{d\theta^2}{\theta^2}\frac{\alpha_s(\theta\pt R)}{\pi}\:\int_{z_\text{min}}^{z_\text{max}} dz\,p^{\text{(finite)}}_{ij}(z),
\end{equation}
where the integration boundaries $z_\text{min}$ and $z_\text{max}$
depend on which matrix element we consider and should match those
imposed by the mMDT conditions in
Eq.~(\ref{eqn:sudakov-integral-representation}).
Once again, to our accuracy, there is some freedom in the choice of the
upper integration boundary of the $\theta^2$ integration. Setting it
to $\zc$ ensures that there are no corrections beyond the transition
point $\rho=\zc$.
Note that neglecting the finite $\zc$ effects is equivalent to keeping
only the contribution from $\mathcal{R}_i$ while neglecting the
contribution from Eq.~(\ref{eq:result-finite}).

\subsection{Impact of the $z$ factor in the scale of the running
  coupling}\label{app:zinalphas}
\begin{figure}[t]
  \centering
   \includegraphics[width=0.495\textwidth,page=1]{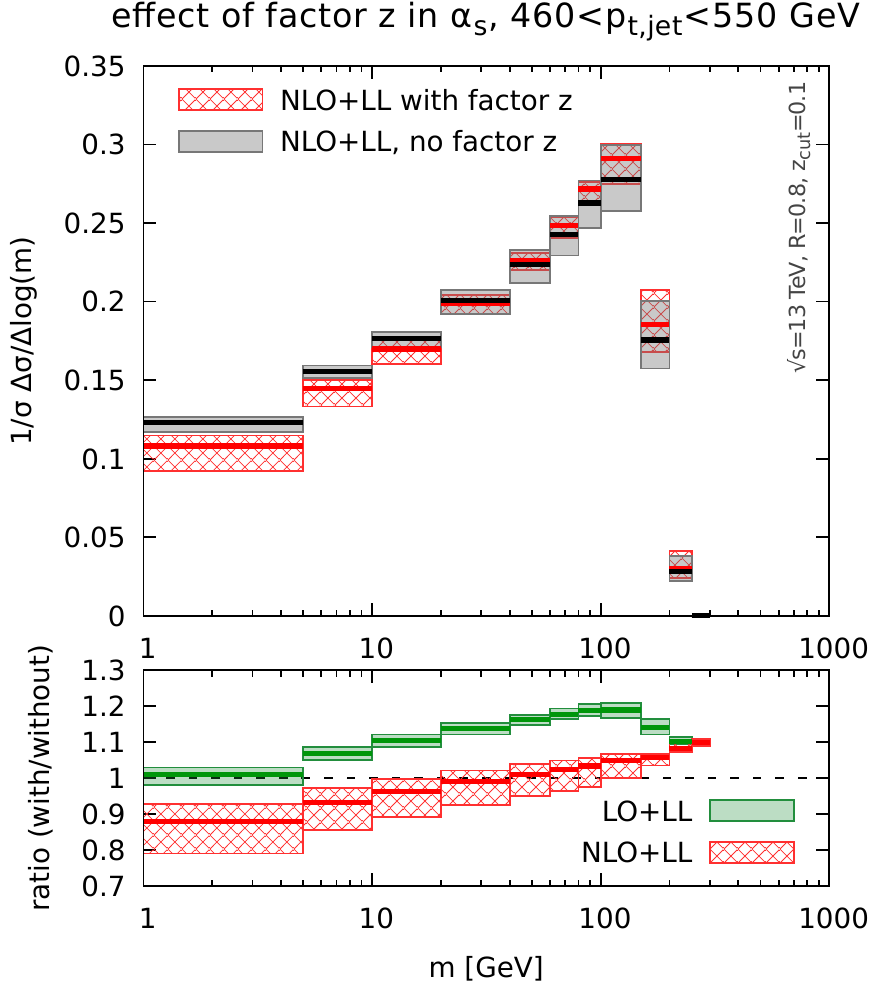}
   \includegraphics[width=0.495\textwidth,page=2]{figs-paper/ptjet-matched-resumz.pdf}
   \caption{Comparison of the jet mass distribution with and without
     the resummation of logarithmic corrections in $\zc$ originating
     from the running of the strong coupling. We note that these
     effects are sizeable, although still within the theoretical
     uncertainty.}\label{fig:logzc}
\end{figure}

If the parameter $\zc$ is chosen to be rather small, finite-$\zc$
corrections are negligible but logarithmic corrections can become
relevant. The resummation of the leading-logarithmic corrections in
$\zc$ is relatively straightforward and it was discussed in
Ref.~\cite{Dasgupta:2013ihk} (see also
Refs.~\cite{Frye:2016okc,Frye:2016aiz}).
Firstly, successive gluon emissions must be ordered in mass rather
than in angle. Secondly, the argument of the QCD
running coupling should be taken as $k_t =z \theta \pt R$ (at least
for the calculation of $\mathcal{R}_i$).
Both effects are included in our calculation.
However, to LL accuracy (in $\rho)$ the argument of the running
coupling could more simply be chosen as $\theta \pt R$. This choice
leads to simpler analytic expressions and is what we naturally obtain
when we consider bins of $\ptg$, see Eq.~(\ref{t-def}).
It is therefore of some interest to investigate how neglecting the
factor $z$ in the argument of the running coupling affects our
findings.
In this case, the $\mathcal{R}_i$ functions in Eq.~(\ref{eq:Rsoft-and-I}) become 
\begin{equation}\label{eq:result-1-no-z}
\widetilde{\mathcal{R}}_i=\frac{1}{\pi\alpha_s\beta_0^2} \Big[
   W\big(1+\alpha_s\beta_0(\log(\rho)-B_i)\big)
   -W\big(1+\alpha_s\beta_0\log(\rho/z_m)\big)
   -\alpha_s\beta_0(\log(z_m)-B_i)\Big].
\end{equation}
In Fig.~\ref{fig:logzc} we show the impact of these corrections on the
normalised matched distributions. Remembering that the uncertainty on
the lower panels is the actual uncertainty on the ratio, we see that
the effects are genuinely present. However, they remain within our
overall theoretical uncertainties shown on the mass distribution (upper plots).

\section{End-point of the $\rho$ distribution}\label{app:rhomax}

As discussed in Section~\ref{sec:match-ptjet}, we have modified the
argument $\log \left( 1 / \rho \right)$ to take into account end-point
effects i.e.\ the fact that $\rho$ has a maximum value $\rho_{\max}$
for a jet with transverse momentum $p_t$ and radius $R$.
In this Appendix, we give the details of the computation of
$\rho_{\max}$ at LO and NLO.

\begin{figure}
  \centering
    \includegraphics[scale=1.0]{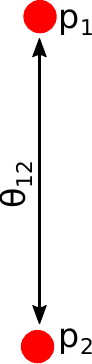} 
    \hspace{2.5cm}
    \includegraphics[scale=1.0]{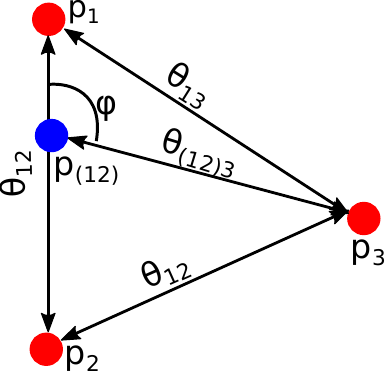}
  \caption{Configurations with maximal mass for LO (left) and NLO (right). }
  \label{fig:diagram-lo}
\end{figure}

At LO, where we have two partons $p_1$ and $p_2$ in the jet, the
calculation is straightforward. The mass of the jet, and therefore
$\rho$, will be maximal when the final partons are as distant as
possible, but are still clustered into a single jet.
Let us first work in the small-angle limit. Then, the angular distance
between the two partons is $\theta_{12} = R$, as shown in the left
plot of Fig.~\ref{fig:diagram-lo}. If the two partons carry a
transverse momentum $p_{t1} = x p_t$ and $p_{t2}=(1-x)p_t$,
respectively, the jet mass is given by
\begin{equation}
m^2 = p_t^2 R^2 x (1-x).
\end{equation}
This is maximal when the momentum is equally distributed between the
two partons, $x=1/2$, for which we have
$\rho^{\text{(small-$R$)}}_{\max,\text{LO}} = 1/4$.
If we relax our small-angle approximation, we should take into account
that the mass of the system of two partons separated by a distance $R$
will depend on their orientation in the rapidity-azimuthal angle
plane. 
It is straightforward to include this in the above analytic
calculation and we find that $\rho$ is maximal when the two partons
have the same rapidity, leading to
$\rho_{\max,\text{LO}} =
R^{-2}\tan^2\frac{R}{2}$~\cite{Dasgupta:2012hg}.
For our choice of $R=0.8$, this gives $\rho_{\max,\text{LO}} = 0.279303$.

At NLO, the same reasoning applies but is complicated by the presence
of one more parton in the jet. We start again by considering the
small-$R$ limit. Remembering that the three partons must be clustered
into a single anti-$k_t$ jet of radius $R$, we can assume, without
loss of generality, that $p_1$ and $p_2$ are the first pair of partons
to be clustered into a subjet with momentum $p_{12}$, with $p_{12}$
then clustered with parton $p_3$.
In order to have all 3 partons clustered into a single jet, we must have $\theta_{12} \le R$ and $\theta_{(12) 3} \le R$.
We define $\varphi$ as being the angle between $\theta_{12}$ and
$\theta_{(12) 3} $, as shown in the right plot of
Fig.~\ref{fig:diagram-lo}, and we parametrise the momentum fractions
of the partons as
\begin{equation}
z_1 = x t, \qquad z_2 = x (1-t), \qquad z_3= 1-x.
\end{equation}
Since $\theta_{(12)1} = (1-t) \theta_{12}$ and  $\theta_{(12)2} = t
\theta_{12}$, we have
\begin{eqnarray}
\theta_{13}^2 &=& (1-t)^2 \theta_{12}^2 + \theta_{(12) 3}^2 + 2 (1-t) \theta_{12} \theta_{(12) 3} \cos \varphi, \\
\theta_{23}^2 &=& t^2 \theta_{12}^2 + \theta_{(12) 3}^2 + 2 t \theta_{12} \theta_{(12) 3}\cos \varphi.
\end{eqnarray}
The jet mass is then found to be
\begin{eqnarray}
m^2 = p_t^2(z_1z_2\theta_{12}^2+z_1z_3\theta_{13}^2+z_2z_3\theta_{23}^2)  =  p_t^2 x t (1-t) \theta_{12}^2 + p_t^2 x (1-x) \theta_{(12)3}^2.
\end{eqnarray}
This is maximal for $\theta_{12} = \theta_{(12)3} = R$ and momentum
equally distributed between $p_1$ and $p_2$, {\it i.e. } $t=1/2$, in
which case we have
\begin{eqnarray}
m^2 =  p_t^2 R^2 x\left( \dfrac{5}{4} - x\right).
\end{eqnarray}
The maximum jet mass is thus reached for $x = 5/8$, which corresponds
to $\rho^{\text{(small-$R$)}}_{\max,\text{NLO}} = 25/64$.
If we lift the small-$R$ approximation, the situation becomes more
complex since the mass now depends explicitly on the angle $\varphi$
as well as on an additional overall rotation angle $\psi$ of the
3-parton system.
One can write analytic expressions for the jet mass and
transverse-momentum conservation and, for given values of $\varphi$
and $\psi$ we can maximise the mass. 
The maximisation over $\varphi$ and $\psi$ has been done numerically
--- imposing that $\Delta R_{12} < R$ and $\Delta R_{(12) 3} < R$ as
required by the clustering --- and we find is $\rho_{\max,\text{NLO}}
= 0.44974$ for $R=0.8$.

\section{LL predictions for the $\ptg$ jet
  cross-section}\label{app:ptmmdt-perturbative}

Before investigating in detail the double-differential cross-section
$d^2\sigma/(d\ptg\,dm)$, one might be tempted to study the jet
cross-section, $d\sigma/d\ptg$.
Despite looking simpler, the latter is actually plagued with the issue
of IRC unsafety, while for the former, the measured jet mass acts as a
regulator of the collinear divergence. 
In this Appendix, we therefore briefly depart from our study of the
double-differential mass distribution to concentrate instead on the
Sudakov-safe $d\sigma/d\ptg$.

\begin{figure}
\centering
\includegraphics[width=0.495\textwidth,page=1]{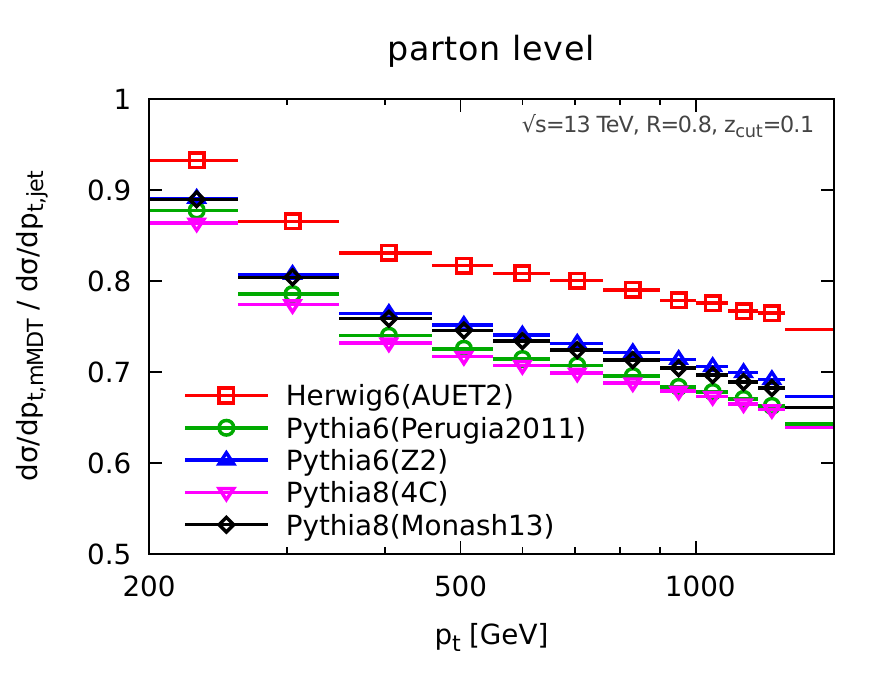}%
\hfill%
\includegraphics[width=0.495\textwidth,page=2]{figs-paper/ptmmdt-ptjet-ratio.pdf}\\
\includegraphics[width=0.495\textwidth,page=3]{figs-paper/ptmmdt-ptjet-ratio.pdf}%
\hfill%
\includegraphics[width=0.495\textwidth,page=4]{figs-paper/ptmmdt-ptjet-ratio.pdf}
\caption{Ratio of the jet cross-section $d\sigma/d\ptg$, binned in the
  groomed jet $p_t$, to the standard jet cross-section
  $d\sigma/d\pt$. The results of Monte-Carlo simulations performed
  with different generators and tunes are shown in the top-left,
  bottom-left and bottom-right plots, respectively for simulations at
  parton-level, hadron-level without UE, and hadron-level including
  UE.
  The top-right plot instead shows our LL analytic results.
}\label{fig:ptmmdt-inclusive-ratio}
\end{figure}

\begin{figure}
\centering
\includegraphics[width=0.52\textwidth,page=5]{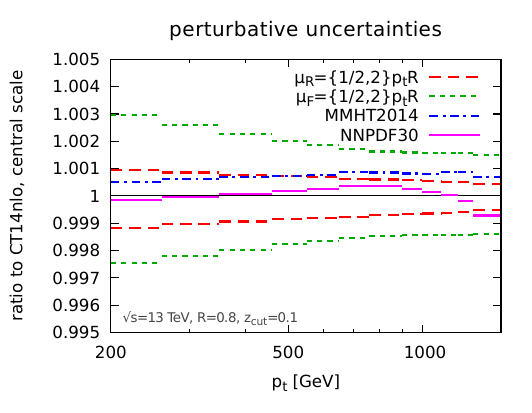}%
\caption{Theoretical uncertainties on the ratio
  $(d\sigma/d\ptg)/(d\sigma/d\pt)$. Uncertainties associated with the
  choice of the renormalisation and factorisation scales as well as
  with the choice of PDF are shown relative to the ratio obtained for
  the central scale choice and our default CT14nlo PDF
  set.}\label{fig:ptmmdt-inclusive-ratio-uncertainties}
\end{figure}

The results of both our LL calculation and of Monte Carlo
simulations at different levels are presented in
Fig.~\ref{fig:ptmmdt-inclusive-ratio}, for the ratio
$(d\sigma/d\ptg)/(d\sigma/d\pt)$.
We can make two main observations: firstly, our LL calculation
provides a good description of what is observed at parton
level. Secondly, as already noticed in
Fig.~\ref{fig:np-effect-inclusive}, hadronisation effects are sizeable
while UE correction are more modest.
Additionally, Fig.~\ref{fig:ptmmdt-inclusive-ratio} shows the
dependence of our LL calculation when varying the value
$t_{\text{max}}$ of $t$ at which we stop parton branchings.
For all the results presented in the main body of the paper, we have
adopted $t_{\text{max}}=1.2$ which shows stable results in
Fig.~\ref{fig:ptmmdt-inclusive-ratio}.

From a theoretical viewpoint, $d\sigma/d\ptg$ can be viewed as the convolution of
the jet spectrum $d^2\sigma/d\pt$ with the ``jet energy drop'',
$1/\sigma\,d\sigma/d\Delta_E$ distribution, computed in the original
Soft Drop paper~\cite{Larkoski:2014wba} at LL accuracy in $\Delta_E$,
neglecting finite $\zc$ corrections.
For the specific case of mMDT, i.e. the limit $\beta\to 0$ of Soft
Drop, we found the remarkable property that, modulo running coupling
corrections, the energy drop spectrum is independent of $\alpha_s$ and
of the flavour of the parton initiating the jet.\footnote{See Eq.~(5.9) of Ref.~\cite{Larkoski:2014wba}.}
It is therefore interesting to study the theoretical uncertainty of
our LL calculation of $d\sigma/d\ptg$, as measured from scale variation. This is shown in
Fig.~\ref{fig:ptmmdt-inclusive-ratio-uncertainties}.
The observed theoretical uncertainty is indeed very small, well below
1\%.
This should be contrasted with the much larger spread of the
parton-level results from our Monte Carlo simulations, the top-left
panel of Fig.~\ref{fig:ptmmdt-inclusive-ratio}. This could be related
to subleading effects not captured by scale variation, or to effects
of finite shower cut-off, seen also in our LL calculation when varying
$t_{\text{max}}$.
The question of the power corrections to the $\ptg$ cross-section, and
to Sudakov-safe observables in general, is therefore interesting both
from the point of view of Monte-Carlo simulations and all-order calculations.

\bibliographystyle{jhep}
\bibliography{biblio}
\end{document}